# All roads lead to (New) Rome:

# Byzantine astronomy and geography in a rapidly changing world

Richard de Grijs[1]

**Abstract.** During the first few centuries CE, the centre of the known world gradually shifted from Alexandria to Constantinople. Combined with a societal shift from pagan beliefs to Christian doctrines, Antiquity gave way to the Byzantine era. While Western Europe entered an extended period of intellectual decline, Constantinople developed into a rich cultural crossroads between East and West. Yet, Byzantine scholarship in astronomy and geography continued to rely heavily on their ancient Greek heritage, and particularly on Ptolemy's *Geography*. Unfortunately, Ptolemy's choices for his geographic coordinate system resulted in inherent and significant distortions of and inaccuracies in maps centred on the Byzantine Empire. This comprehensive review of Byzantine geographic achievements—supported by a review of astronomical developments pertaining to position determination on Earth—aims to demonstrate why and how, when Constantinople fell to the Turks in 1453 and the Ottoman Empire commenced, Byzantine astronomers had become the central axis in an extensive network of Christians, Muslims and Jews. Their influence remained significant well into the Ottoman era, particularly in the context of geographical applications.

**Contents.** 

## 1. Byzantium, intellectual crossroads between East and West

On 11 May 330 CE, Emperor Constantine I the Great (ca. 272–337 CE) ceremonially inaugurated Byzantion (Byzantium) as capital of the Eastern Roman Empire by renaming it Constantinople.[2] The settlement was home to just 5,000 residents in 324 CE, when construction of the new city had commenced. Nevertheless, Byzantion had been inhabited since 658 BCE, when—according to just one of numerous foundation narratives—the legendary, ancient Greek King Byzas of the Megarians established the area's first colony at the mouth of the Bosphorus.[3]

With the Western Roman Empire in a state of terminal decline, Constantinople became a focal point of sorts for scholarly activity, decisively continuing the ancient Greek and Roman traditions. A university-like institution known as the *Pandidacterium* was established in 425 CE by Emperor Theodosius II (401–450 CE). The *Pandidacterium* was, in essence, an educational institution with a number of endowed teaching positions. Despite wars raging in the Eastern Roman Empire—clearly challenges to intellectual pursuits, which were further compounded by natural disasters—education and instruction of the technological and theoretical advances of the

[1] School of Mathematical and Physical Sciences, Macquarie University, Sydney, NSW 2109, Australia.
E-mail: richard.de-grijs@mq.edu.au.
ORCID: 0000-0002-7203-5996.

[2] The city was formally named *N(u)ova Roma* (New Rome) by the Roman Senate.

[3] Alexander A. Vasiliev, *History of the Byzantine Empire, 324–1453* I (Madison: University of Wisconsin Press, 1958): 57–58.

Greeks of the Classical and Hellenistic periods (the last five centuries BCE) continued to be taught,[4] although with an undeniable shift in focus to practical applications.

Ancient Greek manuscripts were readily available to Byzantine scholars, given that they had been collected in Alexandria, far from the strife affecting Rome and its dependencies. Yet, the adoption of Christianity as the Byzantine Empire's official religion, and its focus on teaching theology rather than theory, affected scientific progress and unencumbered intellectual pursuits to varying degrees. Adoption of Christian doctrines initially somewhat stifled further advances in philosophy and, particularly, mathematics.[5] Rhetoric was seen as more important than original thought. As such, philosophical pursuits were reoriented towards finding explanations of natural phenomena, and scientific originality was no longer seen as a key priority.

Nevertheless, the literature, science and philosophy of the ancient Greeks were not completely forgotten. Their prized manuscripts continued to be copied, even under the watchful eye of powerful Christian rulers.[6] Meanwhile, a strong interest in gaining improved practical geographic knowledge was triggered by ecclesiastical, political and commercial drivers, particularly in support of religious pilgrimages to the Holy Land and Jerusalem. As such, astronomy, *logistiki* (accounting, practical arithmetic) and geodesy (also considered a form of arithmetic), in the Aristotelian meaning of 'surveying',[7] retained their strong practical appeal. These subjects were taught both at public universities and by private tutors, first and foremost for practical (usually commercial) use.

According to Hero(n) of Alexandria's (10–75 CE) *Definitiones*,[8] geodesy

> … considers shapes that are not perfect and not clearly defined, … like logistic. Thus, it measures heaps [of earth] as cones, and circular wells as cylinders, and sharp prisms as truncated cones. As geometry employs arithmetic, so this [science] employs logistic. It employs instruments – the *dioptra*[9] for measuring fields, while it uses the rule, the plumb line [level], the gnomon and similar instruments for measuring distances and heights, sometimes with shadows, at others by sightings, and sometimes, it employs light reflection to solve these problems. As the geometer employs imaginary lines, so the geodesist measures perceptible lines. Of these, the most precise are determined by the Sun's rays, either through sightings or by the use of shadows, while the most perceptible are determined by tension and the drawing/pulling of ropes and the plumb line. Using these [instruments], the geodesist measures remote plots of land, the heights of mountains and walls, the width and depth of rivers, and similar things. Geodesy even divides [the Earth], not only into equal parts, but also into secants and proportions and ratios, and often, in accordance with the value of fields.[10]

Dating back to the days of Pythagoras (ca. 570–ca. 495 BCE), the subjects of arithmetic, geometry, astronomy and music became known in Byzantium as the *tetraktys*.[11] As medieval Europe further west descended into an extended period of intellectual decline, Constantinople developed into a rich cultural crossroads between East and West, fostered and fed by the ancient Greek and Roman traditions inherited and retained by contemporary Byzantine practitioners.

Byzantine astronomical and geographic scholarship has long remained relatively poorly studied, although that situation is gradually improving. In this essay, I aim to offer a comprehensive review of Byzantine astronomy from a scientific perspective (rather than from more philosophical or theological angles), particularly in the context of geographic position determination. I start by reviewing, in Section 2, the continuing importance of mostly practical astronomy during the Byzantine era. As we will see, Byzantine astronomy and geography relied heavily on ancient Greek sources, particularly during the first few centuries of the Byzantine Empire. In Section 3, I provide a brief review of those influences in the context of cartographic developments in the Eastern Mediterranean. This is followed by a discussion of Byzantium as the centre of the known world (Section 4) and Constantinople's evolving geographic location as the centuries progress and measurement methods adapt (Section 5). We will return to ancient Greek influences in Section 6, followed by a brief but important detour to influences from further east (Section 7). I then offer my concluding thoughts (Section 8). Ultimately, my aim is to provide a comprehensive framework of Byzantine progress in astronomy and geography—a combintion of topics that is as yet relatively poorly covered in the scholarly literature—while inviting the reader to explore individual aspects in more detail.

---

4 Dimitrios A. Rossikopoulos, "The geodetic sciences in Byzantium", in *Measuring and Mapping the Earth*, eds. Aristeidis Fotiou, Ioannis Paraschakis, Dimitrios Rossikopoulos (Thessaloniki: Special issue for Professor Emeritus Christogeorgis Katsikis, 2015): 1–20, https://www.topo.auth.gr/wp-content/uploads/sites/111/2021/12/01_Rossikopoulos.pdf (accessed 6 November 2023).

5 *Ibid*., 3.

6 Muhammad A. Al-Bakhit, Louis Bazin and Sékéné M. Cissoko (eds.), *History of Humanity, IV: From the Seventh to the Sixteenth Century* (London: Routledge, 1996): chap. 6.

7 Rossikopoulos, "The geodetic sciences in Byzantium", 3, note 3.

8 Knorr (1993) argues convincingly that this treatise should instead be attributed to the Greek mathematician Diophantus of Alexandria (b. ca. 200–214 CE; d. ca. 284–298 CE). Wilbur R. Knorr, "*Arithmêtikê stoicheiôsis*: On Diophantus and Hero of Alexandria", *Historia Mathematica* 20 (1993): 180–192, https://core.ac.uk/download/pdf/82138187.pdf (accessed 6 November 2023).

9 An instrument comparable to the modern theodolite, used for a variety of purposes, including geodesy, astronomy and military applications; see, e.g., Rossikopoulos, "The geodetic sciences in Byzantium".

10 Cited by Rossikopoulos, "The geodetic sciences in Byzantium", 4.

11 In the Western Roman Empire, these subjects were known as the *quadrivium*.

## 2. Astronomy's enduring appeal

During the first few centuries CE, with the centre of the known world gradually shifting from Alexandria to Constantinople, combined with a societal shift from pagan beliefs to Christian doctrines, Antiquity gave way to the Byzantine era. Yet, although ancient Greek treatises—predominantly Claudius Ptolemæus' (Ptolemy; ca. 100–ca. 170 CE) principal manuscripts, the *Almagest*, its derivative *Handy Tables*[12] and the *Planetary Hypotheses*—were routinely copied, extensively discussed and used by experienced practitioners from at least the third century CE, it took until the seventh century CE before the first astronomer of note—Stephanus of Alexandria (*fl.* ca. 580–ca. 640; see below)—made his mark in the Byzantine Empire. Astronomy had, however, become a practical, 'received' science, not a discipline for further exploration and discovery. Byzantine scholars restricted themselves predominantly to commenting on the ancient Greek texts rather than improving or expanding the scholarly knowledge base—a *status quo* that persisted well into the fourteenth century.

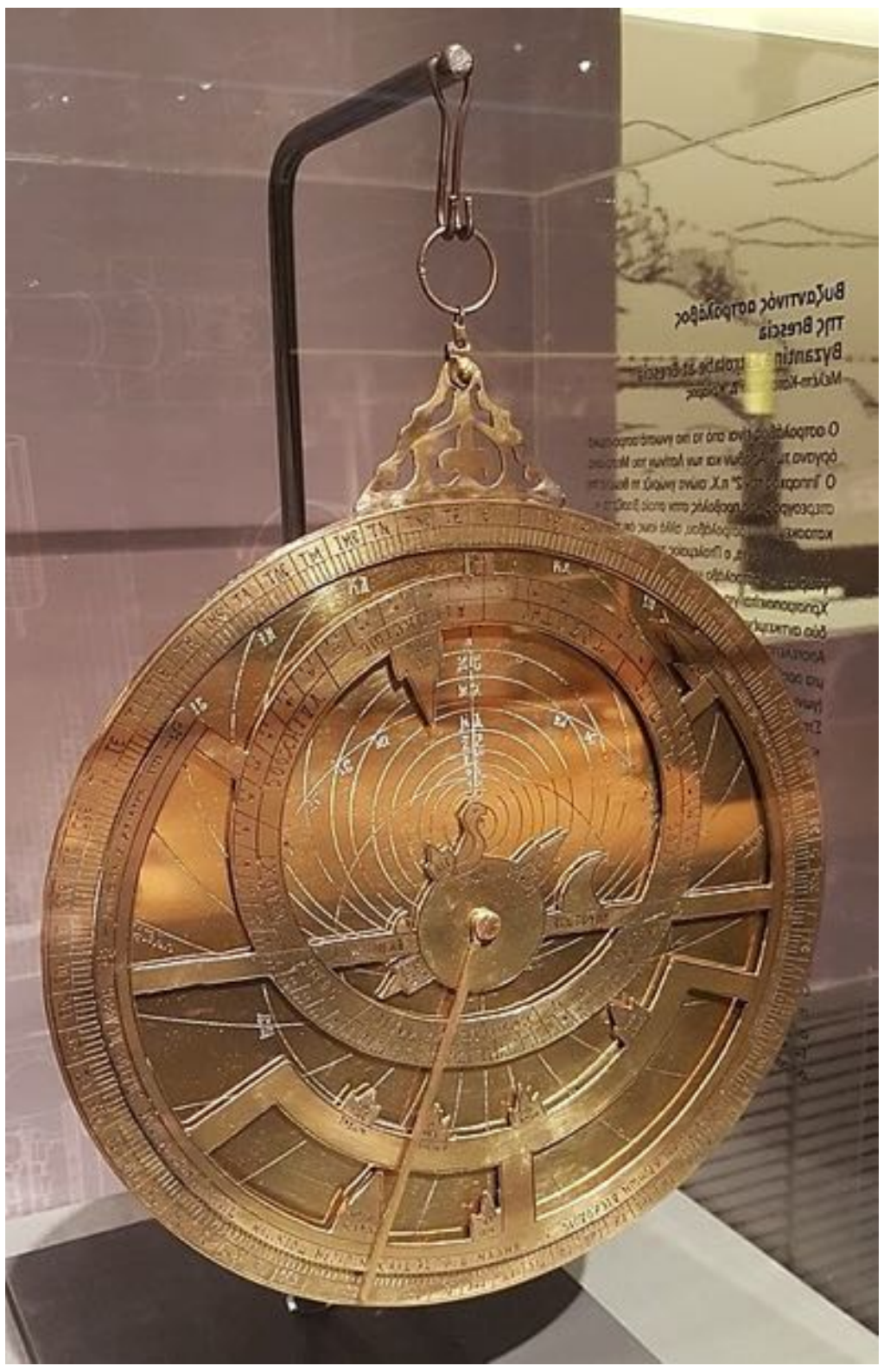

Fig. 1. Reconstruction of Byzantine astrolabe at Brescia (Thessaloniki Technology Museum; credit: Gts-tg via Wikimedia Commons; CC BY-SA 4.0)

Among influential sixth-century philosophers of note, at least in an astronomical context, were John Philoponus (John the Grammarian; ca. 490–ca. 570 CE)—whose commentary on Aristotle's work (fourth century BCE) was influential, although controversial—and Eutocius of Ascalon (ca. 480s–ca. 520s). The latter's commentary on and dissemination of the works of both Archimedes of Syracuse (ca. 287–212 BCE) and Apollonius of Perga (ca. 240–190 BCE) are seen as important contributions to the survival of Greek science during the early Byzantine Empire.[13] Both were associated with the School of Alexandria, having studied under the tutelage of the Neoplatonic philosopher Ammonius Hermiæ (ca. 440–between 517 and 526 CE). Today, only a commentary by Eutocius on the work of Nicomachus of Geresa (ca. 60–ca. 120 CE) and a manuscript on the astrolabe remain of Eutocius' extensive works.

Little is known about the philosopher and teacher Stephanus of Alexandria. During the reign of Emperor Heraclius (ca. 575–641 CE; reigned from 610 CE), he appears to have been summoned by the emperor for the purpose of casting the ruler's horoscope.[14] It has also been suggested that Stephanus may have ghostwritten a rather long-winded practical user guide to Ptolemy's *Handy Tables* that is frequently attributed to the emperor himself.[15]

The increasing dominance of Christianity and its adherence to strict ecclesiastical doctrines, particularly in Western Europe, stifled much of any scientific and intellectual pursuits in the region. These conditions also affected Byzantine scientific developments (although to a more limited extent) until at least the start of the ninth century CE. Nevertheless, across the Greek sphere of influence, particularly at its periphery, cultural and scholarly exchanges between Byzantine astronomers and their counterparts elsewhere continued apace. As a case in point, throughout the first few centuries CE, sections of the *Handy Tables*, augmented by the *Small commentary* attributed to Theon of Alexandria (ca. 335–ca. 405 CE), the Greek

———

12 The *Handy Tables* combine the astronomical tables from the *Almagest*, with later annotations by Theon of Alexandria. They also include instructions for practical use and additions such as a table of the longitudes and latitudes of a number of principal cities, to facilitate conversion with respect to the geographic coordinates of Alexandria. For context, see Emmanuel A. Paschos and Christos Simelidis (eds.), *Introduction to Astronomy by Theodore Metochites* (Singapore: World Scientific, 2015). For the latest insights into astronomical tables used in the Byzantine era, see Efthymios Nicolaidis' contribution to this thematic journal issue.

13 Rossikopoulos, "The geodetic sciences in Byzantium", 7.

14 Keith M. Dickson, "Stephanos of Alexandria (ca 580? – 640? CE)", in *The Encyclopedia of Ancient Natural Scientists: The Greek Tradition and Its Many Heirs*, eds. Paul T. Keyser and Georgia L. Irby-Massie (London: Routledge, 2008): 759–760, https://www.fulmina.org/wp-content/uploads/2018/05/Ancien-natural-scientists-Greek-tradition-and-heirs.pdf (accessed 6 November 2023).

15 Alexander R. Jones, "Later Greek and Byzantine Astronomy", in *Astronomy Before the Telescope*, ed. Christopher B.F. Walker (London: British Museum Press, 1996): 98–109.

scholar and mathematician, were translated into Latin, while the astrologers Stephanus and Theophilus of Edessa (Theophilus ibn Tuma; 695–785 CE) maintained tentative links with Islamic astrologers.[16]

The development of the astrolabe by Byzantine scholars is a rare astronomical highlight during the early centuries of the Byzantine Empire. Arab scholars first became acquainted with the Byzantine astrolabe in the seventh century CE in Harran, Syria, a major centre of Hellenistic knowledge dissemination. The astrolabe played an important role in Islamic contexts. It was applied to determine the ephemerides of the visible planets which, combined with specific pre-calculated tables, allowed Muslim practitioners to determine the times for their five daily prayers.[17] The instrument was additionally used for geodesy (to determine latitudes), thus allowing Islamic astronomers to find the direction to Mecca. Moreover, the astrolabe was useful as a means to measure angles, thus negating the need for a more complicated survey instrument such as a *dioptra*.

Whereas only a single astrolabe from Byzantine times has survived[18] (see Fig. 1),[19] we have access to 15 contemporary manuscripts (most dating from the late Byzantine Empire) describing the instrument. These include two treatises by the astronomer, historian and theologian Theodore Metochites (1270–1332), with one describing its construction and the other its theoretical background. Other manuscripts include one treatise each by Metochites' student, the mapmaker Nicephorus Gregoras (ca. 1295–1360) and Gregoras' student Isaac Argyros (b. ca. 1312). Arab manuscripts from the eighth century onwards imply healthy cross-fertilisation, and in particular significant efforts expended on translations of ancient Greek sources, which were routinely disseminated via Byzantium.[20]

By the twelfth and thirteenth centuries, Byzantine scholars managed to reacquaint themselves with their own Greek heritage thanks to their Arab counterparts and they no longer solely relied on Arab sources (mostly in the form of astronomical tables). The Byzantines themselves shared their knowledge with their colleagues further west, in Europe, founding monasteries and *scriptoria*,[21] workplaces of monastic scribes. A brief handbook on the astrolabe, translated into Greek, is said to have been "… compiled from various methods taken from a Saracen [Islamic] book …",[22] an assertion further supported by the presence of transliterations of Arabic technical vocabulary. By this time, and particularly by the fifteenth century, astronomical knowledge dissemination had come full circle, given that now the Europeans were driving further improvements in the development and practical (nautical) applications of the astrolabe.

Other than these rare astrolabe-related developments, until the ninth century CE the focus of Byzantine scholarship was largely driven by the tenets of Christian doctrines. However, gradually increasing Islamic influences led to a revival of scholars' interest in the remains of ancient literary manuscripts. In turn, that renewed interest prompted a concerted effort at preservation of those manuscripts as durable parchment codices.[23] This period has become associated with the scholarly achievements of Leo the Mathematician (also known as 'the Grammarian' or 'the Philosopher'; ca. 790–after 9 January 869 CE).

Byzantine scholarship reached its pinnacle between the ninth and eleventh centuries. Astronomical manuscripts remaining from that time include four copies of the *Almagest*, four of the *Handy Tables* (including two featuring commentaries by Theon and Pappus of Alexandria; ca. 290–ca. 350 CE), and one on spherical astronomy.[24] However, whereas the period saw a revival of scholarly interest in ancient astronomical and other scientific texts, new and original contributions were few. New contributions from this time mostly represent works of an astrological nature. The era's focus remained firmly on practical astronomical applications (including on their use for measurements of large terrestrial distances: see Fig. 2),[25] as evidenced, for instance, by a surviving example describing the use of the *dioptra* for astronomical purposes:

> Now that we have referred to the terrestrial applications of the *dioptra*, we are ready—though we are on the ground—to attempt to observe the heavens, thanks to the possibilities of the *dioptra*. With it, we can determine the size of the Sun and Moon, and observe the distance between the stars, whether fixed or "wandering" [planets]. … If we wish to observe a longitude or a distance along the zodiac,

---

16 Jones, "Later Greek and Byzantine Astronomy", 104.
17 Rossikopoulos, "The geodetic sciences in Byzantium", 20.
18 The surviving instrument was constructed for a high-ranking Persian official referred to as Sergius, in 1062. It was found in Brescia, Italy. For details, see Ormonde M. Dalton, *The Byzantine Astrolabe at Brescia* (New York: Oxford University Press, 1926).
19 https://commons.wikimedia.org/wiki/Image:Byzantine_astrolabe_at_Brescia,_11th_century_AD_(reconstruction).jpg (accessed 6 November 2023).
20 Jones, "Later Greek and Byzantine Astronomy".
21 Thomas Cahill, *How the Irish Saved Civilization* (New York: Anchor Books, 1995): 193–196; Emmanuel A. Paschos, *Byzantine Astronomy from A.D. 1300* (Fermilab Technical Publications DO-TH 98/18, 1998), https://lss.fnal.gov/archive/other/do-th-98-18.pdf (accessed 6 November 2023).
22 Cited by Jones, "Later Greek and Byzantine Astronomy", 105.
23 Jones, "Later Greek and Byzantine Astronomy", 104.
24 *Ibid*., 105.
25 https://www.researchgate.net/figure/The-first-page-of-a-copy-of-a-manuscript-of-Ptolemys-Cosmographia-Constantinople-end_fig1_356455604

> we do not perform a chance sighting, but we orient the *dioptra* parallel to the meridian; if we are looking for a north–south longitude or *vice versa*, we orient it to the Equator.[26]

The period bracketed by the eleventh and thirteenth centuries represented the onset of a new wave of exploratory studies and astronomical discoveries in the Byzantine Empire. As early as 1032, Islamic influences started to take systematic precedence over the earlier Ptolemaic methods of calculating astronomical phenomena—such as the Sun's orbit projected onto the sky, the mean motions of the Moon and the visible planets, the rate of the Earth's precession, the computation of horoscope components and the prediction of (predominantly solar) eclipses.[27] As regards the latter, an anonymous text from the 1070s is, in fact, a Greek translation of an original manuscript in Arabic by the polymath Muḥammad ibn Mūsā al-Khwārizmī (al-Khwārizmī; ca. 780–ca. 850 CE), which particularly discusses his observations of the solar eclipse of 20 May 1072 using an astrolabe.

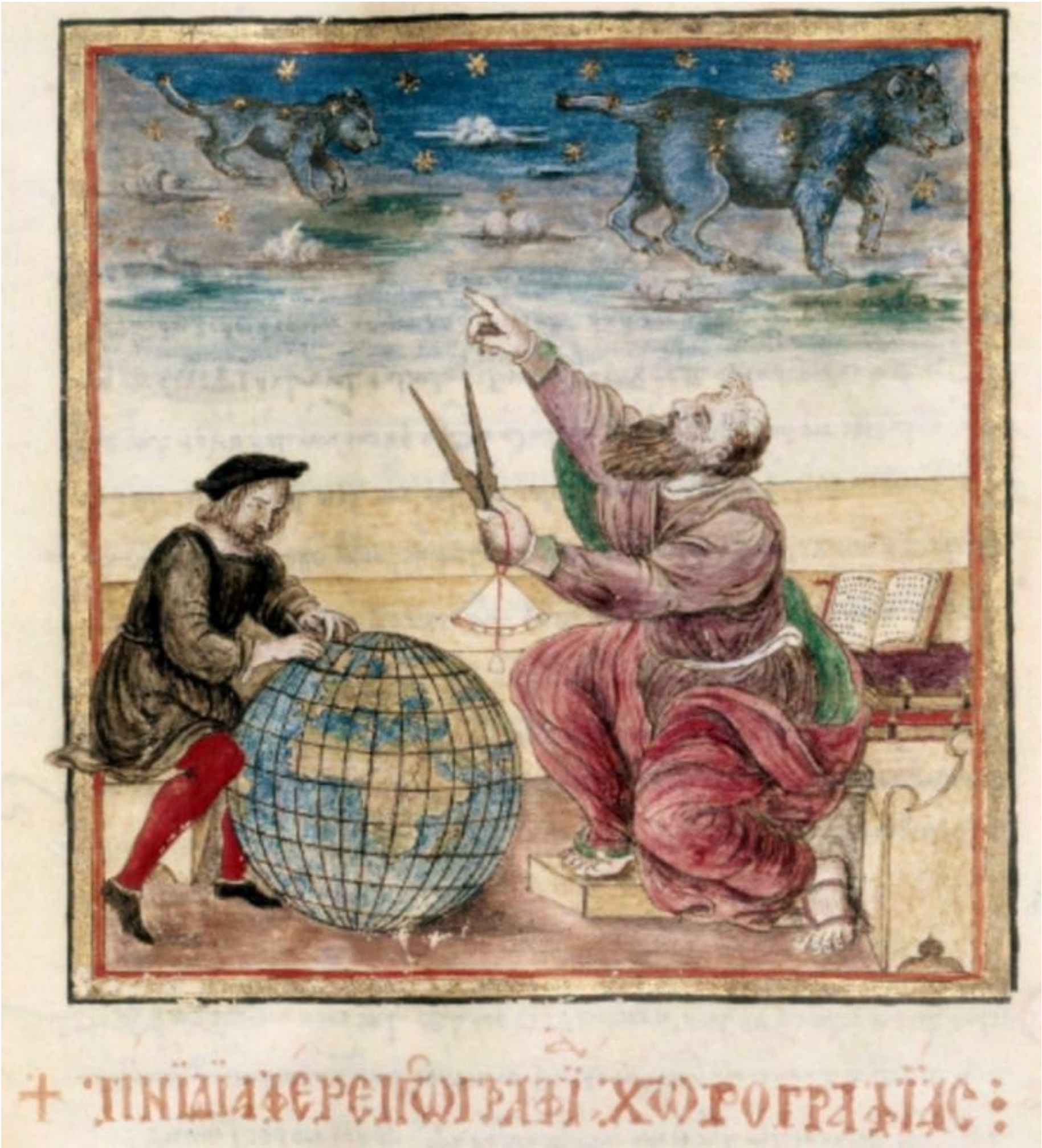

Fig. 2. Cover page of a fifteenth-century Byzantine copy of Ptolemy's *Cosmographia*. Ptolemy himself points to the constellation Ursa Minor with his right hand. With his left hand he is measuring stellar angular separations using a quadrant, a compass and a plumb line. (Bibliothèque Nationale de France: Department of Manuscripts, Codex Greek No. 1401, f. 2; not in copyright)

Byzantine scholars, notably including Gregorios Chioniades (ca. 1240–ca. 1320), hence gradually started to question their received knowledge from the ancient manuscripts and introduced their own improvements.[28] Additionally, stellar catalogues and tables, as well as compilations of the geographic coordinates of the principal cities based on Arab sources, presumably saw regular updates with respect to the ancient Ptolemaic tables. Such sources first allowed Byzantine astronomers in the eleventh century to correct Ptolemy's incorrect latitude determination of Constantinople (see Section 5). Indeed, Jones (1996: 98) enthusiastically muses that

> The Byzantine Greeks['] … reconquest of the theoretical expanses of the *Almagest* after … 1300 was among the intellectual highlights of the Middle Ages, preparing for the developments of the European Renaissance. From Byzantium the Islamic world too drew its knowledge of Ptolemaic astronomy; and in return Byzantine scholars studied and translated Arabic works.

---

26 Anonymous, *Geodesia* (non-contemporary title, 938 CE); cited by Rossikopoulos, "The geodetic sciences in Byzantium", 14.

27 An extant surviving, lengthy commentary on the *Handy Tables* from 1032 relies liberally on Arab sources; Jones, "Later Greek and Byzantine Astronomy", 105.

28 Alexander P. Kazhdan and Ann W. Epstein, *Change in Byzantine Culture in the 11th and the 12th Centuries* (Berkeley/Los Angeles/London: University of California Press, 1990): 140, 149–150, 155; Paschos, *Byzantine Astronomy from A.D. 1300*.

By the thirteenth century, up-to-date—although not necessarily more reliable—astronomical tables were routinely imported from Arab sources and, from the fifteenth century onwards, also from Jewish origins.[29] In turn, Byzantine scholars used their newly found practical astronomical knowledge, alongside the old Ptolemaic tables from the *Almagest* and the *Handy Tables*, to adjust their (Julian) calendar, compute the date of Easter and predict solar and lunar eclipses, among other celestial phenomena. However, rather than using the Arab or Ptolemaic tables in isolation, leading practitioners like the Byzantine monk John Chortasmenos (ca. 1370–before June 1439) often still combined the ancient Greek and contemporary, more user-friendly Persian tables to calculate celestial phenomena.[30]

The Cypriot scholar George Lapithes (*fl.* 1340s), in extensive correspondence with Gregoras, highlighted the lack of precision in contemporary versions of the *Handy Tables* and vowed to work on their improvement with the help of his network of Latin scholars.[31] Lapithes' claim to fame is a 1340s Byzantine adaptation of the *Toledan Tables*.[32] A later Byzantine adaptation of the *Toledan Tables* by Demetrios Chrysoloras (before 1360–after April/May 1416) was based on the longitudes of Cyprus and Paris, implying that the *Tables* reached Constantinople via Cyprus (which was under French rule at the time). Throughout the late Byzantine era, during the period approximately spanning the thirteenth to fifteenth centuries, Byzantine scholars continued to show strong interest in Ptolemy's ancient manuscripts,[33] as evidenced by numerous *scholia*—marginal notes and comments—referring to material from the Greek geographer's *Geographike Hyphegesis* (*Geography*; ca. 150 CE) found in Byzantine copies of treatises by Ptolemy's successor Strabo (64/63–ca. 24 BCE), Plato (428/423–348/347 BCE)[34] and later Gregoras.[35] In Section 3, I will discuss the importance of Ptolemy's *Geography* in the context of Byzantine efforts to quantify the geography of the Eastern Mediterranean.

Perhaps the most original and creative astronomical contribution from Byzantine times was produced by the Neoplatonist philosopher George Gemistos Plethon (ca. 1360–1452). His astronomical tables, calculated for the longitude of Mystras, Greece, were mostly based on Ptolemy's *Almagest*, the *Persian Syntaxis* (the Persian equivalent of the *Handy Tables*) and the Hebrew version of Abū ʻAbd Allāh Muḥammad ibn Jābir ibn Sinān al-Raqqī al-Ḥarrānī aṣ-Ṣābi' al-Battānī's (al-Battānī; before 858–929 CE) astronomical tables known as the *Kitāb az-Zīj aṣ-Ṣābi'* (ca. 900 CE), composed by Immanuel ben Jacob Bonfils (ca. 1300–1377).

Plethon's tables were converted to the longitude of Constantinople as early as 1412–1415, with a revised version appearing by 1446.[36] Plethon's calculations were based on lunar months, thus following the old Greco–Roman calendar, and the start of his 12- or 13-month-long lunisolar year was represented by the first full moon after the winter solstice. This, however, rendered his tables difficult to use by those who based their calculations on the Julian year. Astronomically speaking, Plethon's choices imply that his celestial longitudes used 0° Capricorn as their reference instead of the more common starting point in 0° Aries.[37]

---

29 Alberto Bardi, "Persian Astronomy in the Greek Manuscript *Linköping kl. f. 10*", *Scandinavian Journal of Byzantine and Modern Greek Studies* 4 (2018): 65–88, https://journals.lub.lu.se/sjbmgs/article/view/19699/17804 (accessed 6 November 2023).

30 Anne-Laurence Caudano, "Le calcul de l'éclipse de soleil du 15 avril 1409 à Constantinople par Jean Chortasmenos", *Byzantion* 73.1 (2003): 211–245, https://www.jstor.org/stable/44172822 (accessed 6 November 2023); Bardi, "Persian Astronomy …", 67.

31 Divna Manolova, *Discourses of science and philosophy in the letters of Nikephoros Gregoras* (PhD Thesis, Central European University, 2014): 116–117, https://www.etd.ceu.edu/2015/manolova_divna.pdf (accessed 6 November 2023).

32 During the twelfth century, al-Khwārizmī's tables were gradually replaced by a revised set of tables and an accompanying commentary (the *Canones*), which had been produced specifically for Toledo by the Spanish–Arab (Andalusian) astronomer Abū Ishāq Ibrāhīm al-Zarqālī (1029–1100) in 1080.

33 Anne Tihon, "Les Sciences Exactes à Byzance", *Byzantion* 79 (2009): 380–434, https://www.jstor.org/stable/44173183 (accessed 6 November 2023).

34 Olivier Defaux, "The Iberian Peninsula in Ptolemy's Geography. Origin of the Coordinates and Textual History", *Berlin Studies of the Ancient World* 51 (Berlin: Edition Topoi/Exzellenzcluster Topoi der Freien Universität Berlin und der Humboldt-Universität zu Berlin, 2017): 92, https://doi.org/10.17171/3-51 (accessed 6 November 2023).

35 *India, a Ptolemaic view, c. 200*, https://franpritchett.com/00maplinks/early/ptolemaic/ptolemaic.html (accessed 6 November 2023).

36 Raymond Mercier, "The Astronomical Tables of George Gemistus Plethon", *Journal for the History of Astronomy* 29 (1998): 117–127, https://doi.org/10.1177/002182869802900204 (accessed 6 November 2023); Anne Tihon, "The Astronomy of George Gemistus Plethon", *Journal for the History of Astronomy* 29 (1998): 109–116, https://doi.org/10.1177/002182869802900203 (accessed 6 November 2023).

37 Anne-Laurence Caudano, "Chapter 6. Astronomy and Astrology", in *A Companion to Byzantine Science*, ed. Stavros Lazaris (Leiden: Brill, 2020): 202–230; specifically pp. 226–227.

## 3. Ptolemy and ancient Greek cartographic influences

Much has been written about Ptolemy's *Geography*. I will not regurgitate that discussion here, but I will focus specifically on Ptolemy's choice of coordinate system and its implications for longitude and latitude determinations across *Mare Nostrum* (the Mediterranean), particularly in relation to the inferred geographic location of Byzantium and, later, Constantinople.

### 3.1. Longitudinal distortions across the Mediterranean

Like Marinus of Tyre (ca. 70–130 CE) before him, Ptolemy's world stretched eastwards from the Isles of the Blessed (Blest), also known as the Fortunate Isles, semi-legendary islands in the Atlantic Ocean near the Canary and Cape Verde Islands. However, in his early work, Ptolemy was clearly also influenced by other ancient Greek authors, not least by Eratosthenes of Cyrene (ca. 276–ca. 195/194 BCE). Eratosthenes' work on the 'old map' of Dicæarchus (*fl.* ca. 326–296 BCE)[38] led him to develop a geographical graticule, i.e. a grid of longitudes and latitudes, which he combined with a description of the Earth and its inhabited parts—the *oikumene*: see Fig. 3.[39]

Strabo later commented that Eratosthenes had adopted a meridian that ran through Meroë, Syene (present-day Aswan), Alexandria, Rhodes, Troas (the Hellespont) and the mouth of the Borysthenes (Dnipro River). (Eratosthenes also drew meridians through the Sacred Promontory [Cape St. Vincent], the Strait of Gibraltar, Carthage, the mouth of the Persian Gulf and the delta of the Ganges River.) Following leading thinkers like Hipparchus of Nicæa (190–120 BCE) who had implemented a number of small corrections to Eratosthenes' distance framework and used the 'Alexandrian' axis as the basis of his division of the known world into *climata* (zones of latitude), Ptolemy had adopted this same Alexandrian axis in his *Almagest*. However, by the time he completed his *Geography*, all of the reference locations along the Alexandrian axis had been assigned different longitudes.

Nevertheless, and despite these corrections, Ptolemy's (reconstructed) map of the Eastern Mediterranean still showed the type of distortions in Asia Minor (Anatolia) that are usually associated with the erroneous adoption of the Alexandrian axis:[40] on Ptolemy's map, the longitudes adopted for Alexandria, Rhodes and Byzantium are 60½°, 58⅔° and 56° east of the Fortunate Isles, respectively.[41] A straight line running through all three locations is thus slanted with respect to the usual north–south meridians,[42] placing Byzantium 4½° west of Alexandria and causing distortions in the surrounding areas. In essence, combined with a number of incorrect latitude assumptions (see Section 3.2), adoption of Ptolemy's Alexandrian axis shifts the *Euxine Pontus* (Black Sea) and, hence, the northern coast of Asia Minor westwards by 4½°. A comparison of the locations of Bithynia (northwestern Anatolia) between the *Almagest* and the *Geography*, determined using astronomical means, supports the notion that adoption of the Alexandrian axis introduces systematic distortions.[43]

Recent statistical analysis of some 80 known geographic locations has resulted in the suggestion that the mythical Fortunate Isles may have coincided with the Lesser Antilles, the westernmost islands of the Caribbean;[44] more conventional scholarship places the zero meridian near El Hierro (Ferro), the westernmost of the Canary

---

[38] Strabo, *Geographica* (first century BCE): 2.1.2 (Book 2. Chap. 1. Sec. 2). Edited and translated version: https://www.perseus.tufts.edu/hopper/text?doc=Perseus%3Atext%3A1999.01.0239 (accessed 6 November 2023).

[39] https://commons.wikimedia.org/wiki/File:Mappa_di_Eratostene.jpg (accessed 6 November 2023).

[40] Dmitry A. Shcheglov, "Eratosthenes' Contribution to Ptolemy's Map of the World", *Imago Mundi* 69:2 (2017): 159–175, https://doi.org/10.1080/03085694.2017.1312112 (accessed 6 November 2023).

[41] Strabo reported that Eratosthenes' meridian also ran through the two Cyanean islands in the Black Sea (which remain unidentified today; Lelgemann, 2008) and through the Chalidonia islands near Cape Gelidonya on the southern coast of Asia Minor. Dieter Lelgemann, "On the Geographic Methods of Eratosthenes of Kyrene", in *Proceedings of the International Federation of Surveyors (FIG) Working Week*, https://www.fig.net/resources/proceedings/fig_proceedings/fig2008/papers/hs02/hs02_02_lelgemann_2835.pdf (accessed 6 November 2023).

[42] Modern geographic coordinates place the meridian running through Alexandria some 70 miles (113 km) east of İstanbul.

[43] Francis J. Carmody, "Ptolemy's triangulation of the eastern Mediterranean", *Isis* 67 (1976): 601–609; specifically pp. 608–609, https://www.jstor.org/stable/230563 (accessed 6 November 2023); Elisabeth Rinner, *Zur Genese der Ortskoordinaten Kleinasiens in der Geographie des Klaudios Ptolemaios* (Bern, Tilman Sauer, 2013): 284; Shcheglov, "Eratosthenes' Contribution to Ptolemy's Map of the World", chap. 4.

[44] Liliana Curcio, "Lucio Russo: L'America dimenticata. I rapporti tra le civiltà e un errore di Tolomeo", *Lettera Matematica* 2 (2014): 111–112, https://doi.org/10.1007/s40329-014-0053-1 (accessed 6 November 2023). Note that the resulting length of a stadion, 155.6 m, closely matches other scholars' calculations; cf. Irina Tupikova, Matthias Schemmel and Klaus Geus, *Travelling along the Silk Road: A new interpretation of Ptolemy's coordinates* (Preprint, Max-Planck-Institut für Wissenschaftsgeschichte, 2014), https://www.mpiwg-berlin.mpg.de/Preprints/P465.PDF (accessed 6 November 2023).

Islands. Given the prevailing uncertainties about the precise location of Ptolemy's prime meridian, comparisons with modern geographic location determinations become increasingly more imprecise as one travels across the Mediterranean from west to east. Specifically, the Ptolemaic coordinates of the Canary Islands place these islands on the same meridian but much further south than their actual location. Ptolemy's longitudinal difference with respect to his nearest reference location on the European coast, Cape St. Vincent, the southwesternmost point of Portugal, is 2½° rather than our modern determinations spanning 4°–9°.[45]

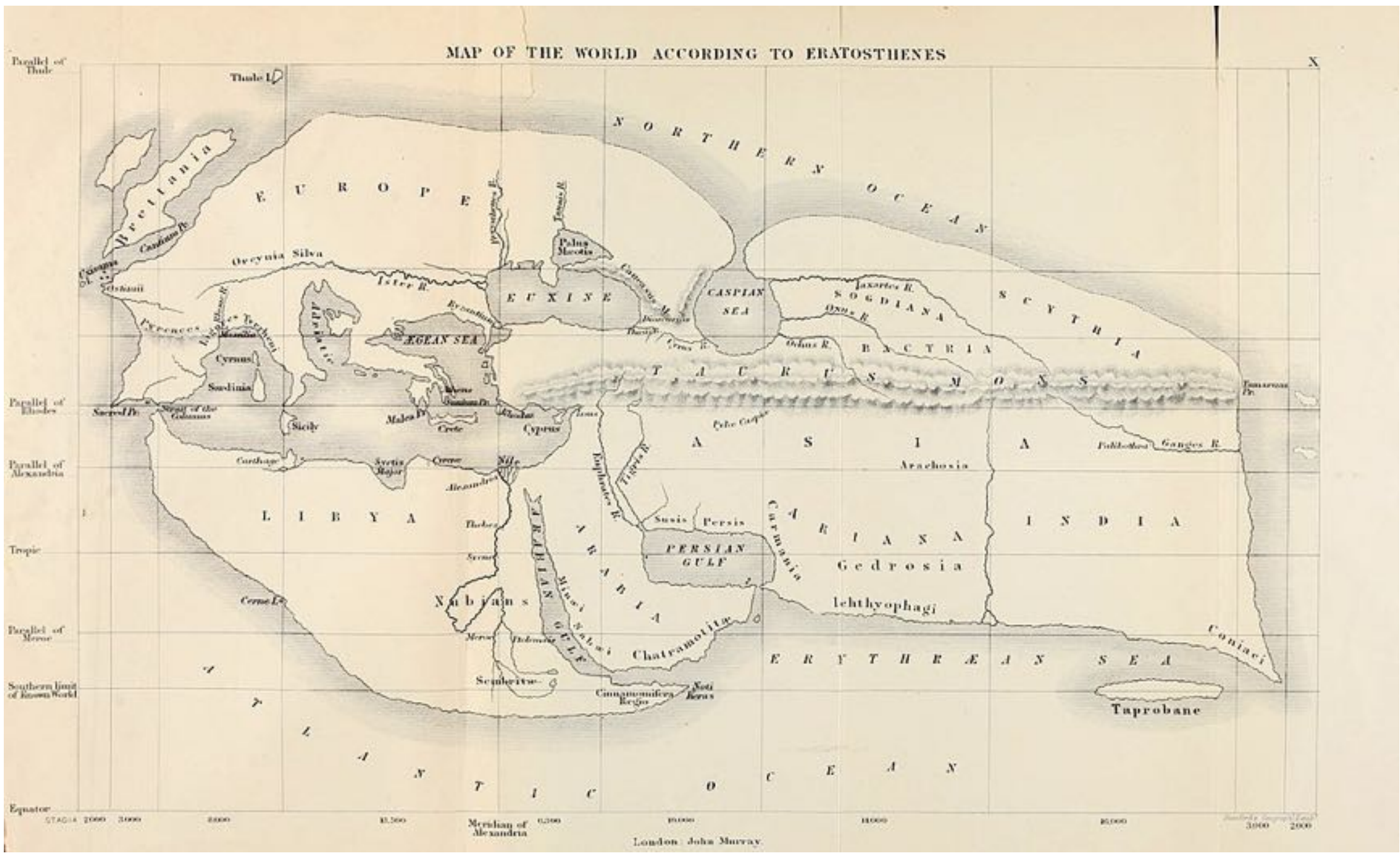

Fig. 3. Nineteenth-century reconstruction of Eratosthenes' map of the known world, ca. 194 BCE.
(Bunbury, 1879: 667; not in copyright)

As one travels further east, the difference in longitude between Ptolemy's values and our modern determinations increases, thus exacerbating the well-known distortions affecting Ptolemy's maps along east–west parallels: the extent of the Mediterranean Sea on Ptolemy's maps is approximately 20° too long. The underlying cause of this east–west distortion has long been known, as we learn from the lucid explanation in Tozer's *A History of Ancient Geography* (1897):[46]

> With regard to the circumference of the Earth Ptolemy followed Marinus in accepting Pos[e]idonius' [ca. 135–ca. 51 BCE] erroneous estimate of 180,000 stadia, which fell short of the reality by one-sixth. It resulted from this that, as he adopted from Hipparchus the division of the Equator and other great circles into 360 degrees, he made every degree only 500 stadia (50 geographical miles) instead of 600 stadia (60 geographical miles), which is the true computation. This mistake at once affected his calculation of distances on his map, for in consequence of it he overestimated them: thus, if he discovered from the authorities – itineraries or otherwise – that the interval between two places was 500 stadia, he would express it on his map by a degree, which in reality is 600 stadia;[47] and then the estimate was made on a large scale, the error in excess became very great. This was especially felt when he came to deal with the second important question of general scientific geography, that of the length of the habitable world, because he greatly overestimated this relatively to the true circumference of the Earth.

However, even after correcting Ptolemy's lineal value of a degree, the extent of Ptolemy's Mediterranean was still about 500 geographical miles too long. It is now commonly accepted that Ptolemy's overestimation of the Mediterranean's extent was caused predominantly by his adoption of the length of the quadrant of a great circle (as on a spherical globe) of 63,000 stadia instead of the more appropriate 54,000 stadia.

---

45 Tupikova *et al.*, "*Travelling along the Silk Road …*", 1.

46 Henry F. Tozer, *A History of Ancient Geography* (Cambridge: Cambridge University Press, 1897): 341–342, https://wellcomecollection.org/works/jxqnp663 (accessed 6 November 2023).

47 Current scholarship implies that one equatorial degree was equivalent to 700 stadia, irrespective of the actual length of a stadion (Tupikova *et al.*, *op. cit.*). This conversion was also adopted by Strabo, following Hipparchus (Bunbury, 1879: 9, note 1). For an in-depth discussion of the mathematical concepts and conversions involved, see Tupikova *et al.* (*op. cit.*). Edward H. Bunbury, *A History of Ancient Geography* II (London: John Murray, 1897), https://archive.org/details/ahistoryancient02bunbgoog/page/n8/mode/2up?view=theater (accessed 6 November 2023).

### 3.2. Inconsistent latitudes and their implications for the location of Byzantium

A second misconception that affected Ptolemy's maps, and which was carried forward well into Byzantine times, was the erroneous acceptance that Massalia (Marseilles) and Byzantium were located on the same parallel (latitude). This notion goes back to Hipparchus. The latter geographer had defined a regular grid pattern of parallels and meridians covering the extent of the *oikumene*. His principal (reference) parallel, located at 36° North (with the Equator serving as zero reference), ran from the Columns of Hercules (the Strait of Gibraltar) through Caralis (Sardinia; 30° 12' N) and Lilybæum (Sicily; 37° 50' N), to the Gulf of Issus (Gulf of İskenderun), passing through Rhodes. As implied by the latitudinal references cited, this adoption led to further distortions in the Eastern Mediterranean; for instance, Carthage was placed 1° 20' south of this parallel, more than 2° south of its actual position.

Hipparchus' system of latitudes is defined by a system of parallels where each parallel's longest day, expressed in solstitial hours (daylight hours at the summer solstice) differs by 15 minutes from that of the next parallel. Following Hipparchus,[48] Ptolemy and Strabo assumed that the longest day in Byzantium lasted 15¼ hours. In Ptolemy's *Almagest*,[49] Book 8 of the *Geography* and the *Handy Tables*, the 14$^{th}$ parallel is assigned a longest day of 15¼ hours; it is specifically associated with Massalia.

Hipparchus drew his next parallel through Byzantium and his birthplace, Nicæa. We learn from Strabo[50] that Hipparchus' conclusion as regards the latitudinal match of Massalia and Byzantium was based on measurements of the ratio of a gnomon to its shadow ($41^4/_5$ : 120) made by Pytheas of Massalia (b. ca. 350 BCE; *fl.* 320–306 BCE) at the summer solstice (at Massalia). Hipparchus is said to have "… found the same relation of the gnomon to its shadow at Byzantium that Pytheas had done at Massilia" [*sic*].[51] Latitude determinations using gnomons should be done on the longest day of the year, at summer solstice, and Ptolemy was clearly aware of that requirement. Such measurements had long been obtained for the principal locations across the *oikumene*, including Alexandria, Babylon, Byzantium and Carthage. However, such measurements often also incurred significant errors, which propagated through contemporary sources until long after Ptolemy's time.[52]

As implied above, Strabo's phrasing of his language implies that Hipparchus may have undertaken such measurements himself, with identical results to Pytheas. In his *Geographica* (first century CE), Strabo states that Hipparchus asserted that the ratio measured by Pytheas in Massalia was the same as that at Byzantium measured at the same time of the year. The measured ratio corresponded to a latitude of 43° 18';[53] the correct latitude is 41° 1'. This displacement of Byzantium by more than 2° northwards also displaced the location of the Black Sea and surrounding areas by the same amount, corresponding to an offset of more than 100 miles (160 km).[54]

As a consequence of Ptolemy's erroneous latitude determinations, the implied north–south extent of the Mediterranean from Massalia to the location directly south of Massalia on the North African coast was 11° instead of the correct extent of 6.5°. Moreover, in Asia Minor, among other distortions, the *Propontis* (the Sea of Marmara) runs north–south on Ptolemy's maps rather than east–west, as in reality, whereas the north and south coasts of Asia Minor have been compressed.[55]

## 4. Centre of an empire

In the year 20 BCE, the Roman Emperor Caesar Augustus (63 BCE–14 CE; born Gaius Octavius) erected the *Milliarium Aureum* (Golden Milestone) near the Temple of Saturn on ancient Rome's *Forum Romanum*. The emperor thus intended to cement Rome as the centre of the known world, with all major Roman roads emanating from the Golden Milestone.

---

48 Rinner, *Zur Genese der Ortskoordinaten Kleinasiens …*
49 *Almagest*, 2.6.14.
50 Strabo, 1.4.4, 2.1.16, 2.5.8, 2.5.40 and 2.5.41.
51 Cited by Bunbury, *A History of Ancient Geography*, 8–9, note 8.
52 Gerd Graßhoff, Florian Mittenhuber and Elisabeth Rinner, "Of paths and places: the origin of Ptolemy's Geography", *Archive for History of Exact Sciences* 71 (2017): 483–508, https://doi.org/10.1007/s00407-017-0194-7 (accessed 6 November 2023).
53 Arabic translations of the *Geography* mention 45° instead; the English mathematician and astronomer John Greaves (1602–1652) determined 41° 6': John Greaves, "An account of the Latitude of Constantinople, and Rhodes, Written by the Learned Mr. John Greaves, sometime Professor of Astronomy in the University of Oxford, and directed to the most Reverend James Usher, Arch-Bishop of Armagh", *Philosophical Transactions of the Royal Society* 15 (1685): 1295–1300, https://doi.org/10.1098/rstl.1685.0093 (accessed 6 November 2023); Graßhoff et al., "Of paths and places: the origin of Ptolemy's Geography", chap. 3.2.
54 Graßhoff et al., "Of paths and places: the origin of Ptolemy's Geography", chap. 3.2.
55 For details, see Rinner, *Zur Genese der Ortskoordinaten Kleinasiens* …; Graßhoff et al., "Of paths and places: the origin of Ptolemy's Geography".

In the fourth century CE, when Constantinople (New Rome) had become the *de facto* capital of the Eastern Romans, Emperor Constantine I the Great ordered the erection of a similar zero-mile marker in the centre of the eastern Empire's principal city. As a result, the 'Stone of Million' or 'Million Stone' (*Milyon Taşı*) was established at the entrance of the Basilica Cistern, today located some 150 m from the Hagia Sophia (whose construction commenced later, around 346 CE) in İstanbul's Sultanahmet Square.

The original monument consisted of a stone in the form of a rectangular prism, some 4–5 m high, covered by a dome located on top of four columns which, in turn, represented four arches, each 14.6 m wide and facing in the four cardinal compass directions (see Fig. 4, right). The arches were crowned by statues of Constantine and his mother, Flavia Julia Helena (Helena of Constantinople; ca. 246/8–ca. 330 CE). Other statues were positioned nearby, whereas the stone itself was richly decorated with reliefs and paintings. Around the sixteenth century, as the Ottoman Empire (1453–1922) took hold, the Million Stone disappeared. Today, all that is left following extensive excavations is the stone itself, as well as the remains of one of the columns, although in badly damaged condition (see Fig. 4, left).

Originally, the base of the Million Stone contained inscriptions of the distances to all of the main cities in the Byzantine Empire: "The stone showed how far other European cities were from Constantinople. The stone featured a plate showing destinations to all the great cities of the time."[56] As such, it acted as the starting point of all main roads across the empire, allegedly giving rise to the expression that "All roads lead to Rome"—New Rome is implied.

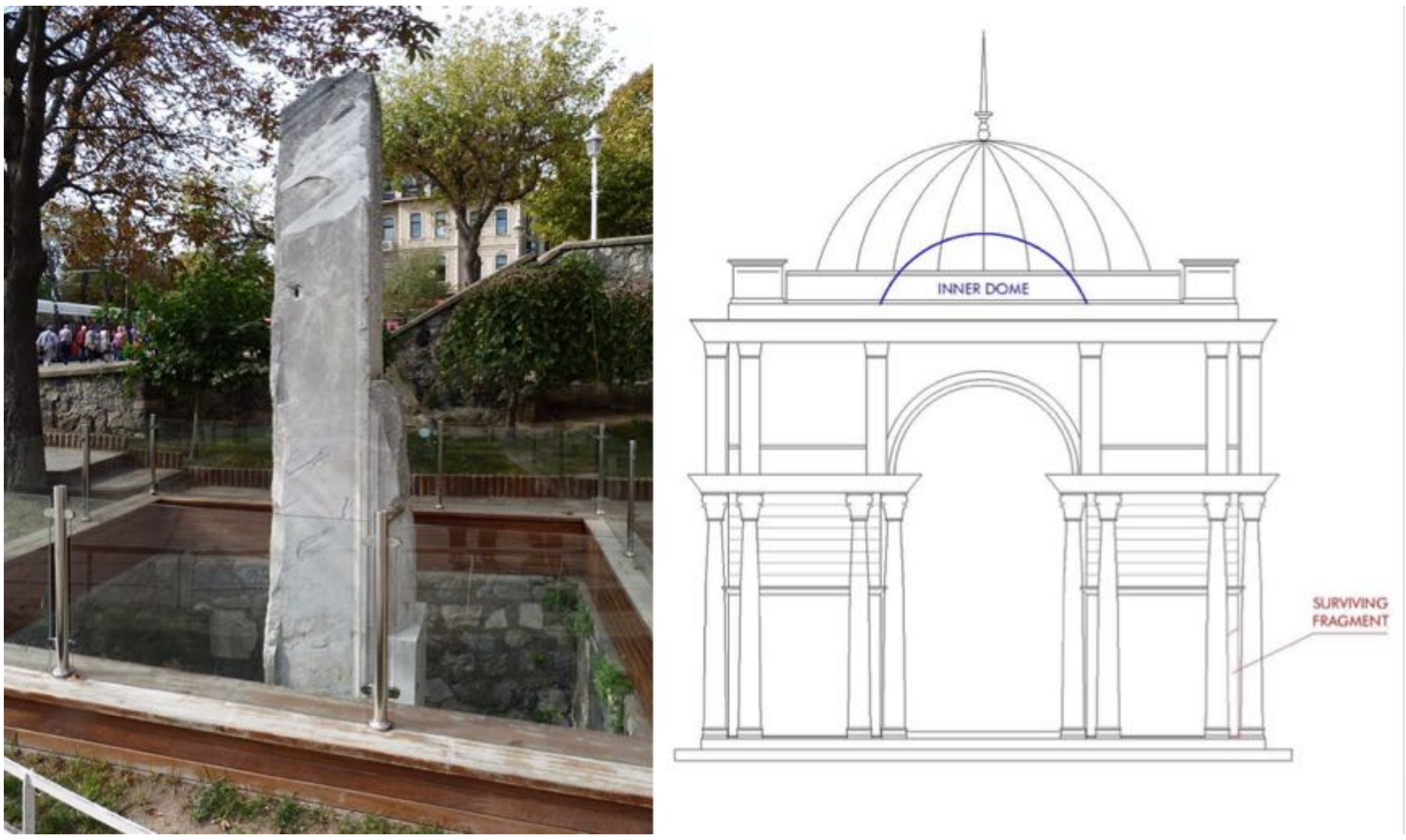

Fig. 4. (left) Badly damaged remains of a pillar that once supported the arches covering the Million Stone in Byzantium. (Credit: Derzsi Elekes Andor via Wikimedia Commons;[57] CC BY-SA 4.0)
(right) Reconstruction of the Million Stone monument (Credit: Greek Strategos via Wikimedia Commons;[58] CC BY-SA 4.0)

Constantinople's dominance eventually placed it at the centre of the Western world, a suggestion further established by Napoléon Bonaparte's (1769–1821) famous exclamation, "If the Earth were a single state, Constantinople would be its capital."[59] And so by the middle of the fifteenth century, from the inception of the Ottoman Empire, one of the world's dominant prime meridians passed through Constantinople. Until the official adoption of the Greenwich meridian as the world's prime meridian in 1884 by the International Meridian Conference in Washington, D.C.,[60] numerous countries consequently adjusted their time keeping by reference to the zero meridian running through Constantinople.

[56] Haldun Hürel, cited by Miraç Tapan, "Journey to the center of the world: Is it far?", *Daily Sabah*, 25 September 2018, https://www.dailysabah.com/feature/2018/09/25/journey-to-the-center-of-the-world-is-it-far (accessed 6 November 2023).
[57] https://commons.wikimedia.org/wiki/File:Milion_Stone_-_Isztanbul,_2014.10.23.JPG (accessed 6 November 2023).
[58] https://en.wikipedia.org/wiki/File:Milion_of_Constantinople_reconstruction.pdf (accessed 6 November 2023).
[59] Napoleon Bonaparte and Jules Bertaut (ed.), "Virilités maximes et ensées" (Nouvelle bibliothèque de variétés littéraires, 1912); cited by Garson O'Toole, http://lists.project-wombat.org/pipermail/project-wombat-project-wombat.org/2014-August/011381.html (accessed 6 November 2023).

## 5. Constantinople's shifting geographic coordinates

Throughout history, Byzantine scholarly interests of scientists, clerics and philosophers alike included applied mathematics and mathematical geography, as well as the shape—particularly the Earth's curvature—and the extent of the seven *climata* and the *oikumene*. The latter was thought to occupy between one-fifth and almost all of the Earth's surface area. Since at least the time of Aristotle the learned and literate members of Greek and Byzantine society had generally accepted that the Earth was a sphere, with few exceptions even among Orthodox Christian theologians.[61] Islamic scholars followed suit.

Dissidents' views supporting a flat Earth were in the minority as early as the third century BCE. Such views, even if expressed as a possibility, did not have any significant effect, not even among religious leaders. As a case in point, Saint John of Damascus (Ioannes Damascenus; ca. 675/6–749 CE) considered both viewpoints, concluding that the shape of the Earth was not a matter of God's creation but for mankind to explore:[62] "Some say that [the Earth] was set and secured on water, like David; others that it [was set] on air."

Byzantine geography relied heavily on Ptolemy's *Geography* and its associated maps, which were reproduced throughout Byzantine history in eight 'Books'. In addition, Byzantine geographers retained the ancient Greek tradition of publishing *periploes*, manuscripts that recorded distances between cities, ports, islands and other important landmarks. One of the earliest-known *periploes*, the *Periplous of Foreign Seas* by the Greek geographer Marcian of Heraclea (*fl.* fourth century CE) explicitly refers to Ptolemy as its source. These references are among the earliest indications underscoring the availability of Ptolemy's *Geography* in Constantinople during Byzantine times.

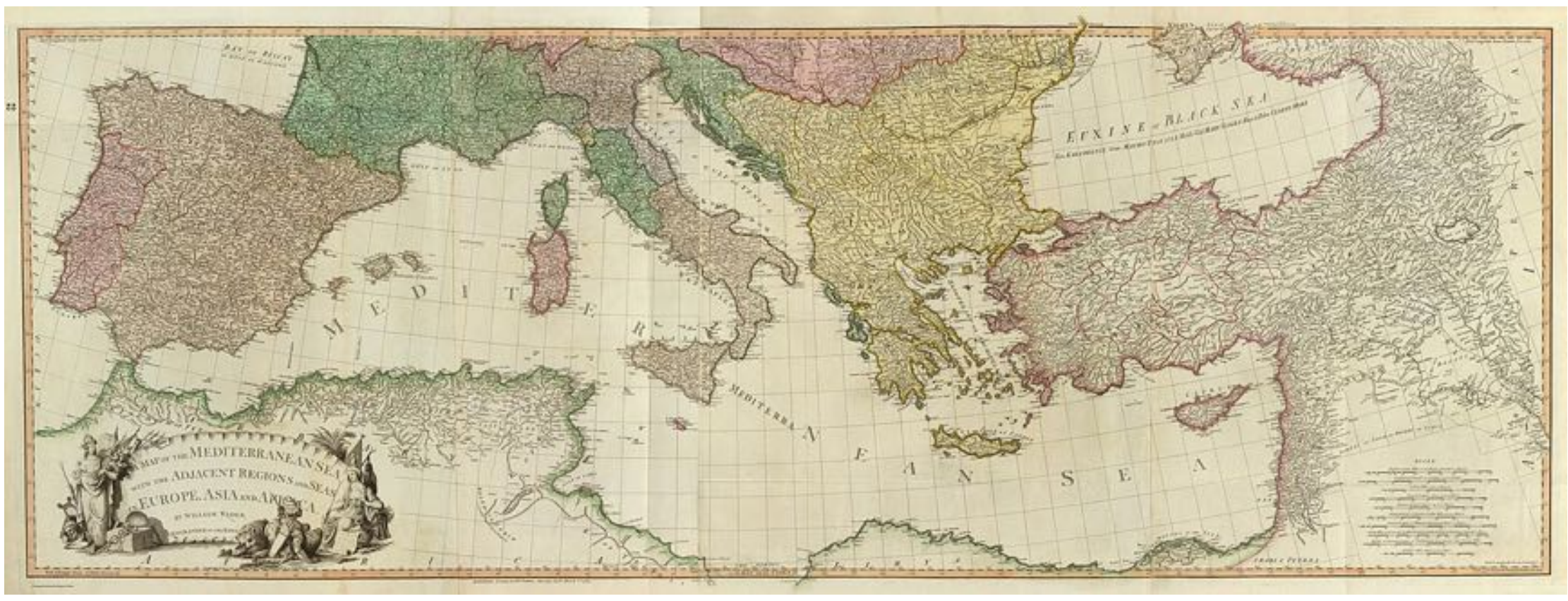

Fig. 5. "*A map of the Mediterranean Sea with the adjacent regions and seas in Europe, Asia and Africa*. By William Faden, Geographer to the King. London, printed for Wm. Faden, Charing Cross, March 1st, 1785." (public domain)

The *Periplous of the Euxine Sea* can be dated to the second half of the sixth century CE. An anonymous copy titled *A Brief Reckoning of the Entire Inhabited World* states in its introduction that the Earth's circumference is 252,000 stadia, with the inhabited region spanning 83,000 × 3500 stadia.[63] The most important Byzantine manuscript of this type is *Stadiasmos*, or *Peripl[o]us of the Great Sea*, which Rossikopoulos (2015) points out, "… describes the *periplous* of the Mediterranean coast, enriched with information concerning distances between ports, descriptions of these, the locations of reefs and shoals, places to obtain supplies, etc." (Standard Byzantine geographic reference points for local maps and charts of the Eastern Mediterranean included Rhodes, Delos and other Aegean islands.) This type of manuscript closely resembles the *portolan* charts that started to appear in the Western Mediterranean by 1300.[64]

---

[60] Richard de Grijs, *Time and Time Again. Determination of Longitude at Sea in the 17th Century* (Bristol: Institute of Physics Publishing, 2017): 7.3–7.5, https://iopscience.iop.org/book/mono/978-0-7503-1194-6 (accessed 6 November 2023).

[61] For a discussion, see Rossikopoulos (*op. cit.*, 6), who points out that the Church Fathers whose beliefs were closer to Platonic philosophy supported a spherical shape, whereas their counterparts from Syria, who tended to follow the teachings of the Old Testament more closely, were more inclined to think of the Earth as flat. See also Jeffrey B. Russel, *Inventing the Flat Earth* (New York: Praeger Press), https://konstantinus.com/wp-content/uploads/2017/01/Russell_Flat_Earth.pdf (accessed 6 November 2023).

[62] Rossikopoulos, "The geodetic sciences in Byzantium", 7.

[63] *Ibid.*, 21.

In the eleventh century, the Greek Byzantine monk Michael Psellos (ca. 1017/8–1078 or 1096) produced an interesting treatise, *De omnifaria doctrina*, describing the Earth as the centre of the universe.[65] The Earth was said to have a circumference of 250,000 stadia, and the elongated *oikumene* covered most of the Earth's surface. Psellos based his insights on Aristotle's *Meteorologica*. The evidence he cites for the Earth's sphericity includes the changing times of sunrise throughout the year, the occurrence of solar and lunar eclipses, the gradual disappearance of ships with increasing distance towards the horizon and the observation that the night sky looks different depending on where one is located on Earth.[66] In the waning days of the Byzantine Empire, Nicephorus Blemmydes (1197–1272) composed two important manuscripts, a *Synoptic Geography* and *Another History of the World in Brief for an Orthodox King*. The latter treatise included a discussion of the Earth's size and curvature and of the seven *climata*.

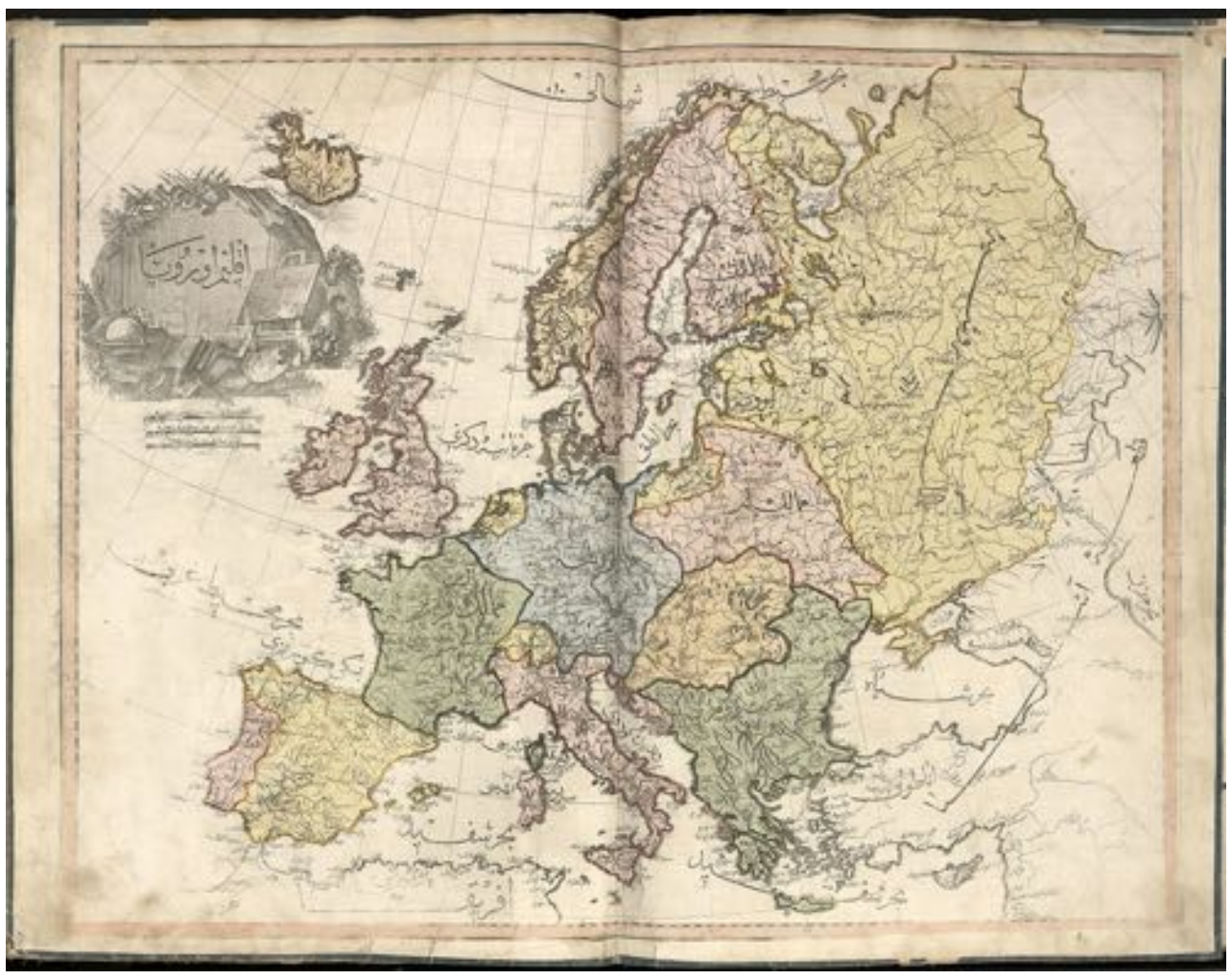

Fig. 6. *Cedid Atlas Tercümesi* (A Translation of a New Atlas) (LOC 2004626120-48; public domain)

Few Byzantine maps remain today, and most of those originate from the thirteenth and fourteenth centuries. By and large, their geography is based on the teachings of Ptolemy and Strabo. The Mediterranean basin was usually adopted as the centre of the known world, with Alexandria—or, more specifically, the Canobic mouth of the Nile, a location in the eastern suburbs of present-day Alexandria—the geographic reference location. The locations of other places were routinely referred to in terms of equinoctial hours from Alexandria.[67] Byzantine geographers were clearly familiar with distances expressed in units of equinoctial hours: for instance, in the sixth century CE, Emperor Justin II (reigned 565–578 CE) presented to the Basilica of Constantinople a *horologium* (a timepiece) beating the hours from one to twelve.[68] However, Byzantine authors and mapmakers also combined the ancient Greek sources with their knowledge of the Bible and information from other, older sources. For instance, Emperor Constantine VII Porphyrogenitus (905–959 CE; reigned from 913 CE) enhanced contemporary information with data from much older sources.[69]

Ptolemy, and subsequently Strabo, had adopted Ferro as his prime meridian. With respect to Ferro, Constantinople was geographically located at the 47th meridian. This standard remained unchanged in Ottoman Türkiye until well into the nineteenth century. Perhaps the most important regional map at this time was William Faden's (1749–1836) *A map of the Mediterranean Sea, with the adjacent regions and seas in Europe, Asia, and Africa* (London, 1785; see Fig. 5),[70] which also adopted Ferro as its prime meridian—despite the prevailing practice to adopt the maker's national capital as their zero-point reference. Faden was the British King George III's (reigned 1760–1820) Royal Cartographer, and so it could have been expected that he would use the Royal Observatory at Greenwich as his geographic reference instead. Faden's influential map, which was brought to Constantinople by Mahmud Râif Efendi (1760–1807), was translated into Ottoman Turkish in 1803, when it became known as the *Cedid Atlas Tercümesi* (A Translation of a New Atlas);[71] see Fig. 6.[72] As a result, the Greek Orthodox cleric Constantius I of Constantinople (1777–1859), who served as Patriarch (Pope) from 1830 to 1834, wrote in his history of the city, *Old and New Constantinople* (1824), "The city is at the 41st latitude and 47th longitude …"

---

64 Frederic C. Lane, "The economic meaning of the invention of the compass", *American Historical Review* 68 (1963): 605–617; specifically pp. 615–616, https://www.jstor.org/stable/1847032 (accessed 6 November 2023).

65 Rossikopoulos, "The geodetic sciences in Byzantium", 10–11.

66 *Ibid*., 11.

67 Paschos and Simelidis, *Introduction to Astronomy by Theodore Metochites*, 367.

68 Charles L. Petze, Jr., "The Evolution of Celestial Navigation. 14. The Chronometer", *Motor Boating* July 1947, 49–51.

69 Al-Bakhit *et al*., *History of Humanity,* chap. 9.3.

70 https://commons.wikimedia.org/wiki/File:William_Faden._Composite_Mediterranean._1785.jpg (accessed 6 November 2023).

71 Emir Alışık, *Eis ten Polin, but Where is the City: Konstantios' Istanbul* (İstanbul Research Institute *Koleksiyon* blog, 11 May 2020), https://blog.iae.org.tr/en/koleksiyon-en/eis-ten-polin-but-where-is-the-city-konstantios-istanbul (accessed 6 November 2023).

72 https://commons.wikimedia.org/wiki/File:Cedid_Atlas_(Europe)_1803.jpg (accessed 6 November 2023).

When the 1884 International Meridian Conference officially adopted the Greenwich meridian as the world's prime meridian, Constantinople's location—and specifically the centre of the dome of the Hagia Sophia—was redefined to align with the 28th meridian, at 28º 58' 50.8188" East of Greenwich, and at a latitude of 39º 36' 00" North.[73]

## 6. Planudes' discovery

Ancient Greek astronomical and geographic achievements saw renewed scholarly interest in the late thirteenth century. At the Greek Orthodox Monastery of Chora,[74] in Constantinople, the Byzantine monk–scholar Maximus Plan(o)udes (ca. 1255–1305) was keenly interested in reviving the ancient geographic manuscripts of Strabo[75] and, particularly, in finding Ptolemy's *Geography*. In 1295, Planudes' efforts paid off: in a poem written in hexameter, he writes enthusiastically about his (re-)discovery and subsequent purchase of a beautiful ancient copy of Ptolemy's *Geography*, lamenting that "… such a great work had been hidden for innumerable years …".[76]

However, one should note that in the Byzantine Empire Ptolemy's *Geography* continued to have an outsized impact on contemporary scholars and never quite faded into oblivion—perhaps because Ptolemy had ensured that his treatises could be used as stand-alone texts without the need to consult other manuscripts. As a case in point, a letter from Constantinople written in 1135 or 1140 from Michael Italicus (*fl.* 1130–1157), a Byzantine medical instructor, to the Byzantine Greek author Theodore Prodromos (ca. 1100–ca. 1165/70) liberally quotes from the *Geography*.[77] Similarly, John Tzetzes (ca. 1120–ca. 1185), the Byzantine poet and grammarian, is known to also have had access to Ptolemy's masterpiece.[78]

Historiographical scholars continue to debate whether the manuscript unearthed by Planudes actually contained any maps.[79] Although Planudes refers to topographical details in his poem,[80] which would suggest that he had access to a cartographic representation of the material contained in his copy of the *Geography*, he also expresses his disappointment at their omission. Nevertheless, his references to visual clues and vivid colours[81] in the first half of the poem have been interpreted as evidence that Planudes was, in fact, in the possession of the type of maps associated with Ptolemy's work. Berggren and Jones (2000: 49–50), however, categorically dismiss the idea that Planudes' rediscovered manuscript might have contained maps:

> The transmission of Ptolemy's text certainly passed through a stage when the manuscripts were too small to contain the maps. Planudes and his assistants therefore probably had no pictorial models … The copies of the maps in later manuscripts and printed editions of the *Geography* were reproduced from Planudes' reconstructions.

This is a common attitude among present-day scholars, which is based on the prevailing assumption that all maps from Ptolemy's *Geography* extant today are Byzantine reconstructions going back to Planudes. This debate continues unabated until the present time. Note, however, that elsewhere in his poem Planudes himself provides us with details about his intent to reconstruct Ptolemy's maps, and how he managed to do so based on Agathodæmon of Alexandria's (ca. second century CE) description of the construction of the *oikumene* and using the coordinates contained within the manuscript's eight Books. Planudes' map (see Fig. 7)[82] reportedly stretched to a length of 17 feet (5.2 m), a suggestion supported by the similarly sized copies found elsewhere (e.g., *Codex Vaticanus Urbinas Grœcus* 82: see Fig. 8).[83]

---

[73] For a technical discussion, see Clifford J. Mugnier, "Republic of Turkey", *Photogrammetric Engineering & Remote Sensing* 82:9 (2016): 671–674, https://www.asprs.org/wp-content/uploads/2018/03/09-16-GD-Turkey.pdf (accessed 10 November 2023).

[74] Based on an *ex libris* in Codex V (presently in the Vatican Apostolic Library and known as *Vaticanus Græcus* 177): '*Claudii Ptolemei liber Geographie et est proprius domini maximi philosophi greci ac monachi in monacerio Chore in Constantinupli. Emptus a quodam Andronico Yneote.*' (Claudius Ptolemy's book *Geography* is the property of the greatest Greek philosopher and monk in the monastery of Chore in Constantinople. Bought from a certain Andronicus Yneote); Florian Mittenhuber, "The Tradition of Texts and Maps in Ptolemy's *Geography*", in *Ptolemy in Perspective: Use and Criticism of his Work from Antiquity to the Nineteenth Century*, ed. Alexander Jones, *Archimedes* 23 (2010): 95–119; specifically p. 119, note 56, https://doi.org/10.1007/978-90-481-2788-7_4 (accessed 6 November 2023).

[75] Rossikopoulos, "The geodetic sciences in Byzantium", 18.

[76] Alfred Stückelberger, "Planudes und die *Geographia* des Ptolemaios", *Museum Helveticum* 53 (1996): 179–205.

[77] Paul Gautier (ed.), "Michael Italikos, Lettres et Discours", *Archives de l'Orient Chrétien* 14 (Paris: Institut Français d'Etudes Byzantines, 1972): 99–101.

[78] Patrick Gautier Dalché, *La Géographie de Ptolémée en Occident (IVe–XVIe siècles)* (Turnhout: Brepols, 2009): 80–81.

[79] Defaux, "The Iberian Peninsula in Ptolemy's Geography …", 62–63.

[80] J. Lennart Berggren and Alexander Jones, *Ptolemy's* Geography. *An Annotated Translation of the Theoretical Chapters* (Princeton: Princeton University Press. 2000): 49.

[81] Filippomaria Pontani, "The World on a Fingernail: an Unknown Byzantine Map, Planudes, and Ptolemy", *Traditio* 65 (2010): 177–200, https://www.jstor.org/stable/41417993 (accessed 6 November 2023); Renate Burri, "Some Notes on the Tradition of the Diagrams (and the Maps) in Ptolemy's *Geography*", in *Claudio Ptolomeo, Geografía (Capítulos teóricos)*, ed./transl. R. Ceceña (Mexico City: Facultad de Filosofía y Letras, Universidad Nacional Autónoma de México. 2018): chap. 1. For instance, Planudes compares his maps' colours to Athena's woven dress or a meadow of flowers.

[82] https://commons.wikimedia.org/wiki/File:PtolemyWorldMap.jpg (accessed 6 November 2023).

Planudes' own map is no longer available; only his commentary has survived. However, it has been suggested that a copy of the *Geography* with maps commissioned by Emperor Andronicus III Palæologus (1297–1341), facilitated by Athanasios I of Constantinople (1230–1310), Patriarch of Alexandria—who happened to be in Constantinople at this time, as we learn from another of Planudes' poems[84]—made its way into the Vatican Apostolic Library.[85] A second known copy is held in the Topkapı Palace in İstanbul. Yet another copy of Planudes' *Geography* and maps is preserved, in part, in the Vatopedi Monastery on Mount Athos, Greece. Unfortunately, this latter copy is incomplete since the monks sold some of the individual maps to collectors; a number of those maps can now be accessed at the British Museum in London, while others have found their way to St. Petersburg, Russia.[86]

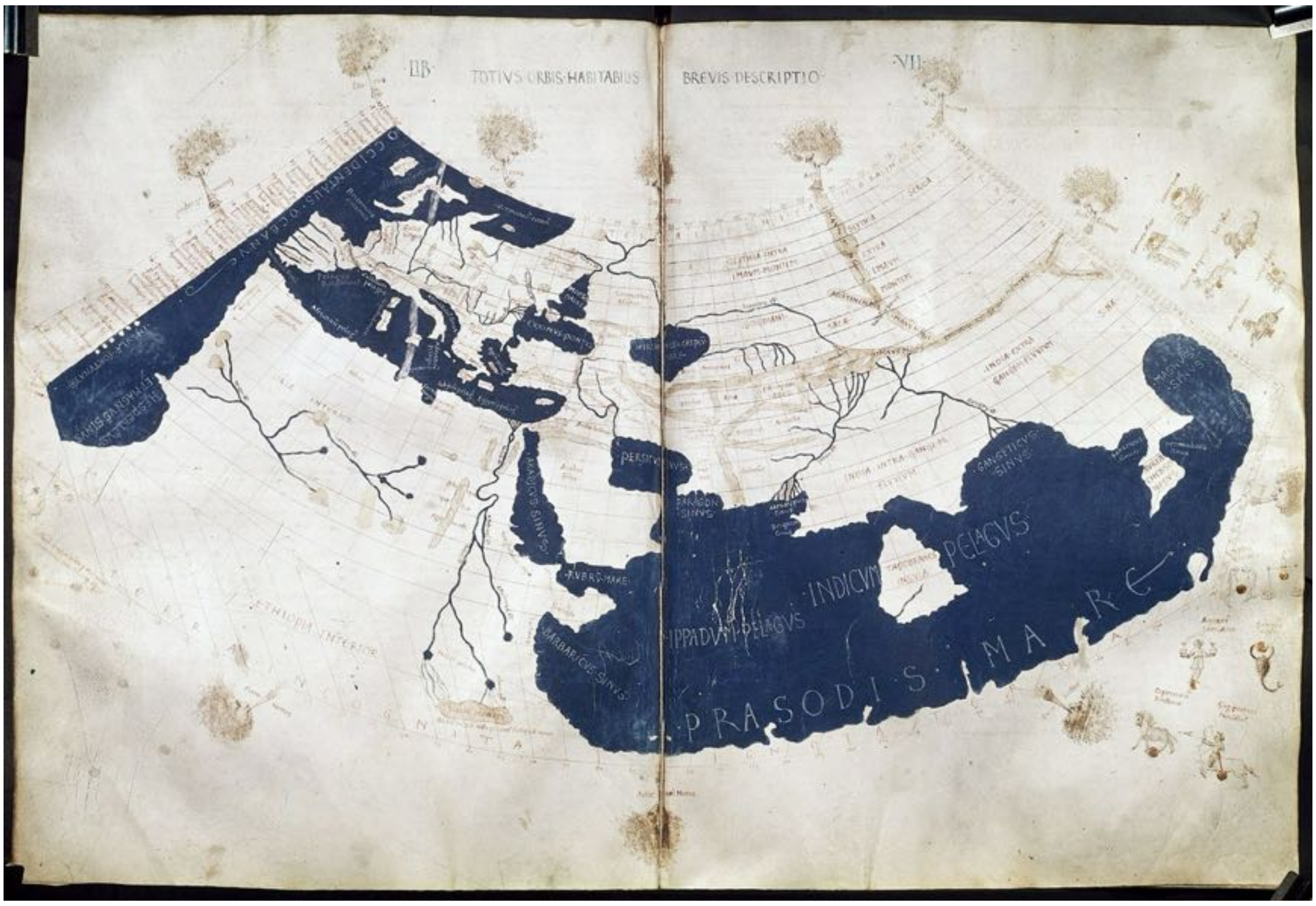

Fig. 7. Ptolemy's world map, reconstituted from the *Geography*; fifteenth-century copy. (Credit: The British Library; public domain)

One might wonder about the provenance of the manuscript version rediscovered by Planudes. Florian Mittenhuber (2010) conjectured that the document may have had its origins in Alexandria, under the stewardship of Athanasios II (d. 496 CE), Patriarch of Alexandria (*fl.* 489–496 CE). He connected the Patriarch's activity with the presence of traces of a Ptolemaic map tradition in Egyptian Islamic culture.

The first contemporary translation into Latin of Planudes' rediscovered version of the *Geography*—complete with maps covering the entire region from the Mediterranean to India—was started by Manuel Chrysoloras (ca. 1350–1415) in 1400. It was completed by one of his students, Giacomo or Jacopo d'Angelo (Jacobus Angelus of Scarparia; 1360–1410), the Italian classical scholar and humanist, in 1406—although d'Angelo confusingly gave it the title *Cosmographia*. The main map in the Chrysoloras/d'Angelo translation includes the location of Byzantium somewhere between the 50$^{th}$ and 55$^{th}$ meridians.[87] This early translation formed the basis for all fifteenth-century renditions of the *Geography*[88] (see, e.g., Fig. 9)[89] and helped popularise Ptolemy's work in

---

[83] Leo Bagrow, *History of Cartography* (London: Routledge, 2009); see also https://digi.vatlib.it/view/MSS_Urb.gr.82. Figure provenance: https://en.wikipedia.org/wiki/File:Ptolemy-World_Vat_Urb_82.jpg (accessed 6 November 2023).
[84] Bagrow, *History of Cartography*, chap. 2.
[85] Rossikopoulos, "The geodetic sciences in Byzantium", 18.
[86] Bagrow, *History of Cartography*, chap. 2.
[87] Alışık, *Eis ten Polin …*
[88] Bagrow, *History of Cartography*, chap. 2.

Western Europe, where few scholars still mastered the Greek language. As we saw at the end of Section 5, Faden's map of 1785 placed Constantinople at the 47th parallel, based on the prime meridian at Ferro.

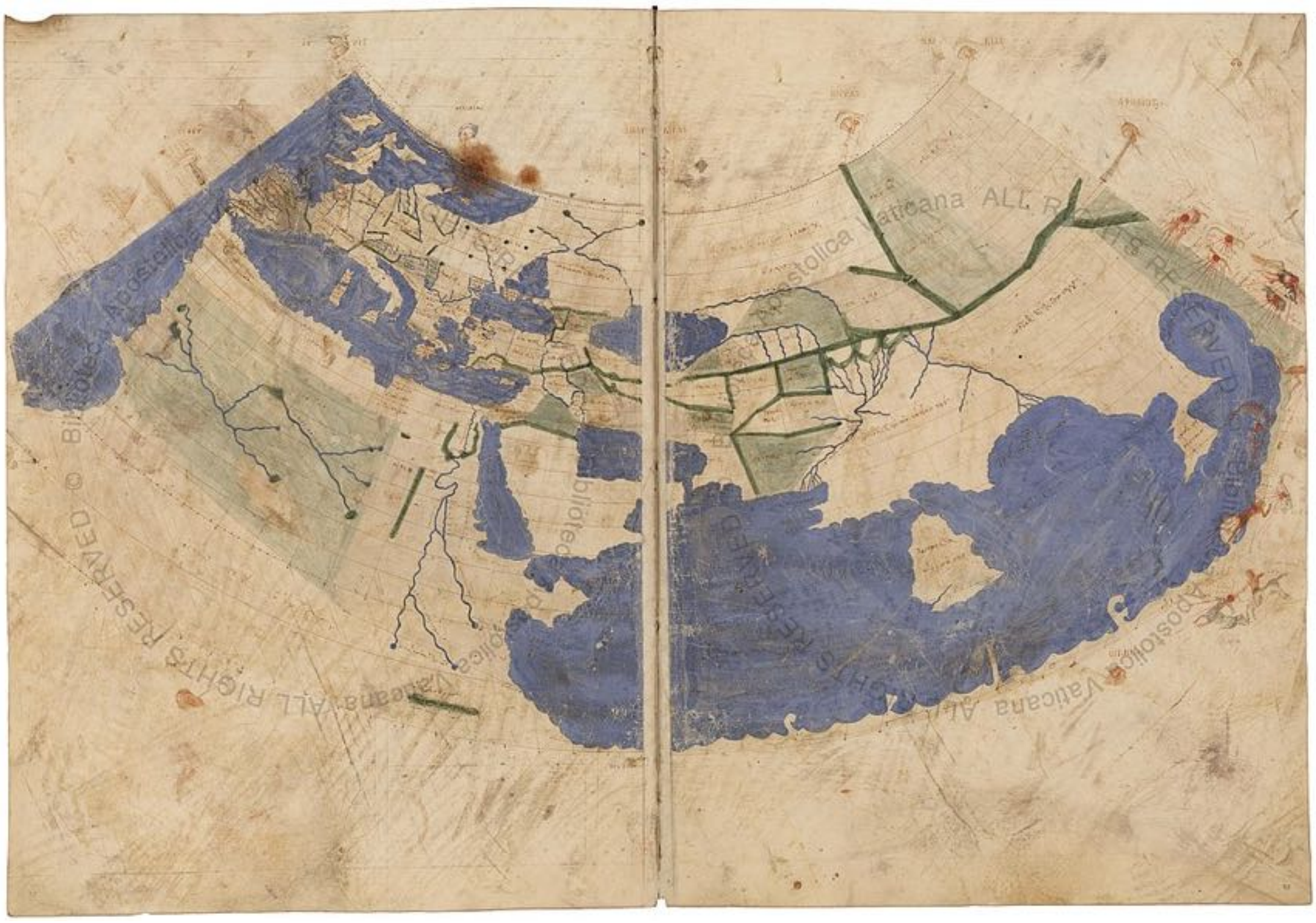

Fig. 8. Byzantine Greek world map from the *Codex Vaticanus Urbinas Graecus* 82; Constantinople, ca. 1300. Parchment 575 × 418 mm. (public domain)

Specific reference is made to the Arab historian and geographer Abū al-Ḥasan ʿAlī ibn al-Ḥusayn ibn ʿAlī al-Masʿūdī's (ca. 896–956) work *Muruj al-dhahab* (Meadows of Gold; ca. 950 CE), which was composed in Egypt. In addition, in an eleventh-century copy of the *Kitab surat al-ard* (Book of the Image of the Earth; 1037), based on al-Khwārizmī's ninth-century original, the representation of the Nile and the known world's *climata* closely follow Ptolemaic tradition, thus firmly connecting geographic scholarship from Antiquity to Byzantine times, clearly facilitated by productive cultural exchanges during the Islamic 'Golden Age' (spanning the eighth to thirteenth centuries).

This cross-fertilisation should not have come as a surprise. Byzantine scholarship continued the long tradition of the study of ancient Greek literature and culture—although by the time Planudes and his contemporaries engaged with Ptolemy's masterwork, Byzantine scholars simply accepted the Greek polymath's work as accurate and indisputable. They did not pursue corrections or updates based on newly obtained geographic insights.

## 7. Growth of Islamic influences

Although early Byzantine science relied heavily on ancient Greek sources, from approximately the ninth century CE, Islamic influences started to appear in Byzantine scholars' dealings with geography and astronomy. This period coincided with the Islamic Golden Age. Islamic geography, while initially still heavily influenced by Ptolemaic traditions (e.g., al-Khwārizmī's ninth-century world map, which adopted a prime meridian coinciding with the Canary Islands), soon developed into a leading specialty on its own accord. Examples of major new developments that were not based on Ptolemaic *climata* include the Balkhī School's *Atlas of Islam*, a set of 20 maps composed by Abū Zayd Ahmed ibn Sahl Balkhī (al-Balkhī ; 850–934 CE) and a set of 70 maps based on the concept of *climata* by Abu Abdullah Muhammad al-Idrīsī al-Qurtubi al-Hasani as-Sabti (al-Idrīsī; 1100–1165), among many others.

Following a peak in Greek scientific achievements in the second century CE, after the fall of the Roman Empire Greek and other Western influences gradually diminished as Western Europe entered an extended period of intellectual decline, an era dominated by myth, superstition and the rigid teachings of Christianity, spanning some six centuries and lasting until the European Renaissance.[90] Greek scientific doctrine and philosophy continued to flourish in Byzantium and the Arab world for some time, but by the ninth century CE developments in Byzantine scientific thinking had similarly slowed down. Once-influential manuscripts—including Ptolemy's *Geography*—became largely lost or forgotten.

[89] https://www.reddit.com/r/MapPorn/comments/7kx0ej/1420_byzantine_copy_of_ptolemys_world_map_4766_x/ (accessed 6 November 2023).

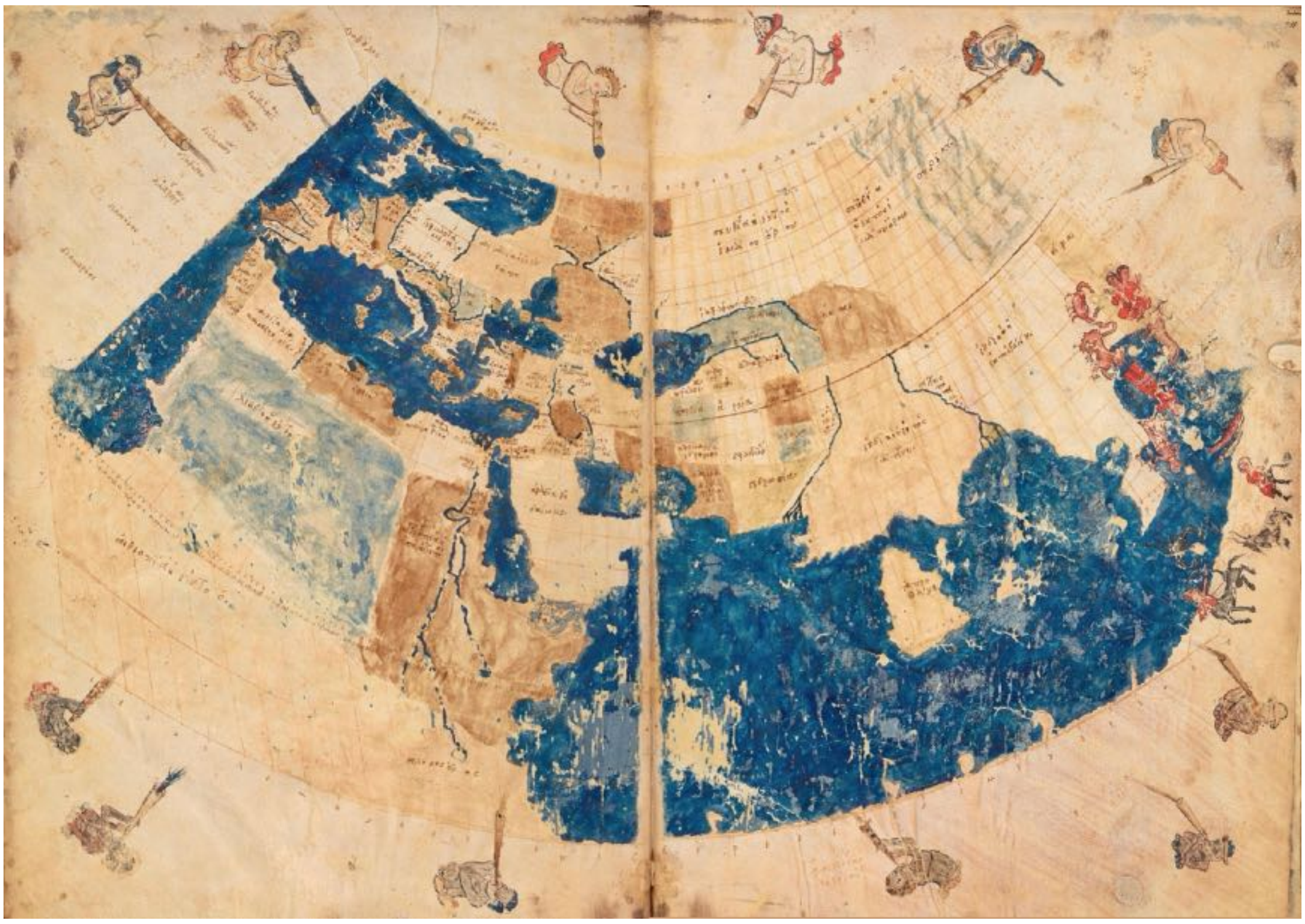

Fig. 9. Byzantine copy of Ptolemy's world map, ca. 1420. (public domain)

Meanwhile, during the eighth and ninth centuries CE, Islamic scholars also expanded their worldview by allowing cross-fertilisation of their work with the scholarly knowledge originating from Greece and Byzantium in the west, and from Persia and India in the east. Greek manuscripts found their way, translated into Arab, to the Islamic world, often via Byzantine interlocutors. The critical assessment of this wealth of new information available to Arab scholars eventually resulted in the Islamic Empire's great prosperity. In turn, this led to further efforts to capitalise on the world's knowledge, with many new translations into Arabic emerging of Byzantine scientific manuscripts.[91]

Yet, by the eleventh to thirteenth centuries, with science, philosophy and scholarship in Byzantium now also in a protracted state of decline, Islamic influences gradually expanded westwards to both Constantinople and the Empire of Trebizond on the southern shores of the Black Sea. A long list of travellers' accounts of Byzantium, the Aral–Caspian region, the empire of the Khazars (nomadic Turkic people) and the 'Rus' state (Russia) contributed significantly to major Islamic progress in cartographic developments well beyond their core region,[92] which kept Byzantine geographic developments alive, albeit at a much lower level than during earlier centuries. This eventually resulted in nautical dominance by Islamic navigators, from the eleventh century onwards, including by such luminaries as Vasco da Gama's (ca. 1460s–1524) pilot, the Arab geographer Ahmad ibn Mājid (ca. 1432–ca. 1500) and Sulaymān ibn Aḥmad ibn Sulaymān al-Mahrī (ca. 1480–1550). Arab nautical maps and charts included full grids of longitudes and latitudes;[93] their navigators used their astronomical knowledge to the fullest extent in support of their sea voyages.

---

90 e.g., Dimitri Gutas, *Greek Thought, Arabic Culture* (London: Routledge, 1998); George Saliba, *Islamic Science and the Making of the European Renaissance* (Cambridge: MIT Press, 2007).

91 *Ibid.*

92 Al-Bakhit *et al.*, *History of Humanity,* chap. 6.

93 de Grijs, *Time and Time Again* …, chap. 2.2.1.

Among the early Arab astronomers whose work had a significant influence on Byzantine astronomy was ʻAlī ibn al-Ḥusayn Abū al-Qāsim al-ʻAlawī al-Sharīf al-Ḥusaynī (also known as Ibn al-Aʻlam; d. 985 CE).[94] Ibn al-Aʻlam, known in Greek as Alim, composed a set of astronomical (planetary) tables in a handbook later referred to as the *Naṣīr al-Dīn al-Ṭūsī's īlkhānī Zīj* (thirteenth century), also known as *al-Zīj al-Sharīf* (named after the author), *al-Zīj al-ʻAḍudī* (named after his patron, ʻAḍud al-Dawla; ruler of Baghdad, 978–983 CE) or *al-Zīj al-Baghdādī*. The reference to Baghdad in this latter title could be a nod to the author's place of residence, or it could imply that the astronomical tables had been composed (most likely from earlier observations[95]) with respect to the Baghdad meridian.

However, unlike similar contemporary handbooks, the *al-Zīj al-Baghdādī* does not contain any tables pertaining to calendars, geographic coordinates, fixed stars or the usual trigonometric and spherical functions. Nevertheless, Ibn al-Aʻlam's work found its way to the Byzantine Empire, given the existence of a version of his tables from 1032 CE, which had most probably been converted to the meridian of Constantinople. There is evidence from Persian and Arab sources that Ibn al-Aʻlam's tables were used for practical astronomy until well into the fourteenth century.[96]

By the fourteenth century, Byzantine astronomy witnessed significant influences from the Islamic world and beyond. The introduction to the *Persian Syntaxis* (1347), which contained astronomical tables of Persian origin edited and translated into Greek by the Greek physician, geographer and astronomer Georgius Chrysococcas (*fl.* 1340s), provided extensive background on scientific exchanges at this time between Persia on the one hand and Byzantium and Trebizond on the other.[97] Chrysococcas's predecessor, the Byzantine astronomer Gregorios Chioniades (ca. 1240–ca. 1320) had acquired his astronomical knowledge in Tabriz (located in present-day Iran) from the Persian scholar Shams al-Dīn Muḥammad ibn Mubārakshāh Mīrak al-Bukhārī al-Harawī (al-Bukhārī; d. 1339). He subsequently brought his knowledge to Trebizond.[98]

Although there is some evidence that Persian astronomical achievements had made their way to Byzantine scholars by the start of the fourteenth century, the earliest translations into Greek of Persian astronomical treatises is usually ascribed to Chioniades. These include the *Zīj as-Sanjarī* (ca. 1120) by Abū al-Fatḥ ʻAbd al-Raḥmān al-Khāzinī (*fl.* 1115–1130), and a commentary by al-Bukhārī on the *Zīj Alā'ī* (ca. 1176) by the Persian astronomer Farīd al-Dīn Abu al-Ḥasan ʻAlī b. al-Fahhād (*fl.* twelfth century).

## 8. Concluding thoughts

As we have seen, during the first few centuries CE, the centre of the known world gradually shifted from Alexandria to Constantinople, with Antiquity giving way to the Byzantine era. And whereas much of western Europe entered an extended period of intellectual decline, Constantinople developed into a rich cultural crossroads between East and West. However, Byzantine scholarship in astronomy and geography was defined by an undeniable shift in focus to practical applications rather than intellectual innovations. Nevertheless, that more practical approach to astronomy and geography may, in fact, have cemented Constantinople's role as the centre of the known world.

Any and all of the scholarly developments discussed in this essay must be considered more broadly in the context of events that happened in Byzantine society. The Byzantine Empire governed Anatolia until they were forced to give up control by the incessant onslaught of Seljuk Turkish forces from the east in 1071 CE. Meanwhile, the Normans, Bulgars and Patzinaks (Petchenegs) attacked the empire from the west. The centre of the commercial world hence gradually shifted to Venice, where conditions were more peaceful. All roads no longer led to New Rome.

Constantinople surrendered to the Crusaders in 1204, but the city was recaptured in 1261. Nevertheless, Muslim Ottoman forces continued to exert pressure during the fourteenth and fifteenth centuries leading to an emigration wave westwards, from Constantinople to Italy. Byzantine Greek manuscripts, including Byzantine

94 Josep Casulleras, "Ibn al‑Aʻlam: ʻAlī ibn al‑Ḥusayn Abū al‑Qāsim al‑ʻAlawī al‑Sharīf al‑Ḥusaynī", in *The Biographical Encyclopedia of Astronomers*, eds. Thomas Hockey *et al.*, Springer Reference (New York: Springer, 2007): 549, https://ismi.mpiwg-berlin.mpg.de/biography/Ibn_al-Alam_BEA.htm (accessed 6 November 2023).

95 *Ibid.*

96 *Ibid.*

97 Raymond Mercier, "The Greek «Persian Syntaxis» and the *Zīj-i Īlkhānī*", *Archives Internationales d'Histoire de Sciences* 112.3 (1984): 35–60; Bardi, "Persian Astronomy …", 66, 70.

98 F. Jamil Ragep, "New Light on Shams: the Islamic Side of Σὰμψ Πουχάρης", in *Politics, Patronage and the Transmission of Knowledge in 13th–15th Century Tabriz*, ed. Judith Pfeiffer (Leiden: Brill, 2014): 231–247.

copies of the *Geography*, served as important collateral.[99,100] Once in Italy, many of those manuscripts were eventually translated into Latin. By 1453, when Constantinople fell to the Turks and the Ottoman Empire commenced, Byzantine astronomers hence formed the central axis in an extensive network of Christians, Muslims and Jews, combining best practices from multiple cultural traditions and facilitating the coexistence of numerous religious communities. Byzantine geograpic influences remained significant well into the Ottoman era. As a case in point, I refer to the geographical teachings of the Greek Renaissance scholar and philosopher George Amiroutzes (1400–1470) in relation to Mehmed II, 'the Conqueror' (1432–1481; reigned 1444–1446 and from 1451).[101]

One of my main aims in composing this essay was a desire to demonstrate and emphasise the importance of cross-cultural communication and the free exchange of ideas on the largest scales, using Byzantine scientific history as my main vehicle. On a related note, geographic innovations during the Islamic Golden Age contributed to significant developments across an enormous swathe of land, from the far western European periphery to Imperial China in the east. However, Islamic scholars also generously borrowed, adapted and adopted ancient Greek and, particularly, Byzantine insights and approaches, thus clealy underscoring the importance of cross-cultural communication and the free exchange of ideas for the benefit of all.

## 9. References

Al-Bakhit, Muhammad A., Louis Bazin and Sékéné M. Cissoko (eds.), *History of Humanity, IV: From the Seventh to the Sixteenth Century*, London: Routledge, 1996.

Alışık, Emir, *Eis ten Polin, but Where is the City: Konstantios' Istanbul*, İstanbul Research Institute *Koleksiyon* blog, 11 May 2020. https://blog.iae.org.tr/en/koleksiyon-en/eis-ten-polin-but-where-is-the-city-konstantios-istanbul (accessed 6 November 2023).

Anonymous, *Geodesia* (non-contemporary title), 938 CE.

Bagrow, Leo, *History of Cartography*, London: Routledge, 2009.

Bardi, Alberto, "Persian Astronomy in the Greek Manuscript *Linköping kl. f. 10*", *Scandinavian Journal of Byzantine and Modern Greek Studies* 4 (2018) 65–88. https://journals.lub.lu.se/sjbmgs/article/view/19699/17804 (accessed 6 November 2023).

Berggren, J. Lennart, and Alexander Jones, *Ptolemy's* Geography. *An Annotated Translation of the Theoretical Chapters*, Princeton: Princeton University Press. 2000.

Bonaparte, Napoléon, and Jules Bertaut (ed.), "Virilités maximes et ensées", Nouvelle bibliothèque de variétés littéraires, 1912.

Bunbury, Edward H., *A History of Ancient Geography* II, London: John Murray, 1897. https://archive.org/details/ahistoryancient02bunbgoog/page/n8/mode/2up?view=theater (accessed 6 November 2023).

Burri, Renate, "Some Notes on the Tradition of the Diagrams (and the Maps) in Ptolemy's *Geography*", in *Claudio Ptolomeo, Geografía (Capítulos teóricos)*, ed./transl. R. Ceceña, Mexico City: Facultad de Filosofía y Letras, Universidad Nacional Autónoma de México. 2018.

Cahill, Thomas, *How the Irish Saved Civilization*, New York: Anchor Books, 1995.

Carmody, Francis J., "Ptolemy's triangulation of the eastern Mediterranean", *Isis* 67 (1976) 601–609. https://www.jstor.org/stable/230563 (accessed 6 November 2023).

Casulleras, Josep, "Ibn al-A'lam: 'Alī ibn al-Ḥusayn Abū al-Qāsim al-'Alawī al-Sharīf al-Ḥusaynī", in *The Biographical Encyclopedia of Astronomers*, eds. Thomas Hockey *et al.*, Springer Reference, New York: Springer, 2007, p. 549. https://ismi.mpiwg-berlin.mpg.de/biography/Ibn_al-Alam_BEA.htm (accessed 6 November 2023).

99 Naftali Kadmon, "Ptolemy, the first toponymist. Submitted by Israel", in 7th Conference of the United Nations Group of Experts on Geographical Names: "*Commemorating the Thirtieth Anniversary of the First United Nations Conference on the Standardization of Geographical Names*". https://unstats.un.org/unsd/geoinfo/ungegn/docs/7th-uncsgn-docs/econf/7th_UNCSGN_econf.91_114.pdf (accessed 6 November 2023).

100 e.g., John E. Sandys, *A History of Classical Scholarship: From the Sixth Century B.C. to the End of the Middle Ages*, 1 (Cambridge, UK: Cambridge University Press, 1903): 376–428; Deno J. Geanakoplos, *Greek Scholars in Venice: Studies in the Dissemination of Greek Learning from Byzantium to Western Europe* (Cambridge: Harvard University Press, 1962); Nigel G. Wilson, *From Byzantium to Italy: Greek Studies in the Italian Renaissance* (Baltimore: Bloomsbury Publishing, 1992); John Monfasani, *Byzantine Scholars in Renaissance Italy: Cardinal Bessarion and Other Émigrés: Selected Essays* (Farnham: Variorum, 1995).

101 See, e.g., references in Brigita Kukjalko, "The Study of Ancient Greek Texts in Early Ottoman Constantinople", *Byzantina Symmeikta* 30 (2020): 283–306; see also Julian Raby, "Mehmed the Conqueror's Greek Scriptorium", *Dumbarton Oaks Papers* 37 (1983): 15–34.

Caudano, Anne-Laurence, "Le calcul de l'éclipse de soleil du 15 avril 1409 à Constantinople par Jean Chortasmenos", *Byzantion* 73.1 (2003) 211–245. https://www.jstor.org/stable/44172822 (accessed 6 November 2023).

Caudano, Anne-Laurence, "Chapter 6. Astronomy and Astrology", in *A Companion to Byzantine Science*, ed. Stavros Lazaris. Leiden: Brill, 2020, pp. 202–230.

Curcio, Liliana, "Lucio Russo: L'America dimenticata. I rapporti tra le civiltà e un errore di Tolomeo", *Lettera Matematica* 2 (2014) 111–112. https://doi.org/10.1007/s40329-014-0053-1 (accessed 6 November 2023).

Dalton, Ormonde M., *The Byzantine Astrolabe at Brescia*, New York: Oxford University Press, 1926.

Defaux, Olivier, "The Iberian Peninsula in Ptolemy's Geography. Origin of the Coordinates and Textual History", *Berlin Studies of the Ancient World* 51. Berlin: Edition Topoi/ Exzellenzcluster Topoi der Freien Universität Berlin und der Humboldt-Universität zu Berlin, 2017. https://doi.org/10.17171/3-51 (accessed 6 November 2023).

de Grijs, Richard, *Time and Time Again. Determination of Longitude at Sea in the 17th Century*, Bristol: Institute of Physics Publishing, 2017. https://iopscience.iop.org/book/mono/978-0-7503-1194-6 (accessed 6 November 2023).

Dickson, Keith M., "Stephanos of Alexandria (ca 580? – 640? CE)", in *The Encyclopedia of Ancient Natural Scientists: The Greek Tradition and Its Many Heirs*, eds. Paul T. Keyser and Georgia L. Irby-Massie, London: Routledge, 2008, pp. 759–760. https://www.fulmina.org/wp-content/uploads/2018/05/Ancien-natural-scientists-Greek-tradition-and-heirs.pdf (accessed 6 November 2023).

Gautier, Paul (ed.), "Michael Italikos, Lettres et Discours", *Archives de l'Orient Chrétien*, 14 (1972) 99–101.

Gautier Dalché, Patrick, *La Géographie de Ptolémée en Occident (IVe–XVIe siècles)*, Turnhout: Brepols, 2009.

Geanakoplos, Deno J., *Greek Scholars in Venice: Studies in the Dissemination of Greek Learning from Byzantium to Western Europe*, Cambridge: Harvard University Press, 1962.

Graßhoff, Gerd, Florian Mittenhuber and Elisabeth Rinner, "Of paths and places: the origin of Ptolemy's Geography", *Archive for History of Exact Sciences* 71 (2017) 483–508. https://doi.org/10.1007/s00407-017-0194-7 (accessed 6 November 2023).

Greaves, John, "An account of the Latitude of Constantinople, and Rhodes, Written by the Learned Mr. John Greaves, sometime Professor of Astronomy in the University of Oxford, and directed to the most Reverend James Usher, Arch-Bishop of Armagh", *Philosophical Transactions of the Royal Society* 15 (1685) 1295–1300. https://doi.org/10.1098/rstl.1685.0093 (accessed 6 November 2023).

Gutas, Dimitri, *Greek Thought, Arabic Culture*, London: Routledge, 1998.

Jones, Alexander R., "Later Greek and Byzantine Astronomy", in *Astronomy Before the Telescope*, ed. Christopher B.F. Walker, London: British Museum Press, 1996, pp. 98–109.

Kadmon, Naftali, "Ptolemy, the first toponymist. Submitted by Israel", in 7th Conference of the United Nations Group of Experts on Geographical Names: "*Commemorating the Thirtieth Anniversary of the First United Nations Conference on the Standardization of Geographical Names*". https://unstats.un.org/unsd/geoinfo/ungegn/docs/7th-uncsgn-docs/econf/7th_UNCSGN_econf.91_114.pdf (accessed 6 November 2023).

Kazhdan, Alexander P., and Ann W. Epstein, *Change in Byzantine Culture in the 11th and the 12th Centuries*, Berkeley/Los Angeles/London: University of California Press, 1990.

Knorr, Wilbur R., "*Arithmêtikê stoicheiôsis*: On Diophantus and Hero of Alexandria", *Historia Mathematica* 20 (1993) 180–192. https://core.ac.uk/download/pdf/82138187.pdf (accessed 6 November 2023).

Kukjalko, Brigita, "The Study of Ancient Greek Texts in Early Ottoman Constantinople", *Byzantina Symmeikta* 30 (2020) 283–306.

Lane, Frederic C., "The economic meaning of the invention of the compass", *American Historical Review* 68 (1963) 605–617. https://www.jstor.org/stable/1847032 (accessed 6 November 2023).

Lelgemann, Dieter, "On the Geographic Methods of Eratosthenes of Kyrene", in *Proceedings of the International Federation of Surveyors (FIG) Working Week.* https://www.fig.net/resources/proceedings/fig_proceedings/fig2008/papers/hs02/hs02_02_lelgemann_2835.pdf (accessed 6 November 2023).

Manolova, Divna, *Discourses of science and philosophy in the letters of Nikephoros Gregoras* (PhD Thesis, Central European University, 2014). https://www.etd.ceu.edu/2015/manolova_divna.pdf (accessed 6 November 2023).

Mercier, Raymond, "The Greek «Persian Syntaxis» and the *Zīj-i Īlkhānī*", *Archives Internationales d'Histoire de Sciences* 112.3 (1984) 35–60.

Mercier, Raymond, "The Astronomical Tables of George Gemistus Plethon", *Journal for the History of Astronomy* 29 (1998) 117–127. https://doi.org/10.1177/002182869802900204 (accessed 6 November 2023).

Mittenhuber, Florian, "The Tradition of Texts and Maps in Ptolemy's *Geography*", in *Ptolemy in Perspective: Use and Criticism of his Work from Antiquity to the Nineteenth Century*, ed. Alexander Jones, *Archimedes* 23 (2010) 95–119. https://doi.org/10.1007/978-90-481-2788-7_4 (accessed 6 November 2023).

Monfasani, John, *Byzantine Scholars in Renaissance Italy: Cardinal Bessarion and Other Émigrés: Selected Essays*, Farnham: Variorum, 1995.

Mugnier, Clifford J., "Republic of Turkey", *Photogrammetric Engineering & Remote Sensing* 82:9 (2016) 671–674. https://www.asprs.org/wp-content/uploads/2018/03/09-16-GD-Turkey.pdf (accessed 10 November 2023).

Paschos, Emmanuel A., *Byzantine Astronomy from A.D. 1300*, Fermilab Technical Publications DO-TH 98/18, 1998. https://lss.fnal.gov/archive/other/do-th-98-18.pdf (accessed 6 November 2023).

Paschos, Emmanuel A., and Christos Simelidis (eds.), *Introduction to Astronomy by Theodore Metochites*, Singapore: World Scientific, 2015.

Petze, Jr., Charles L., "The Evolution of Celestial Navigation. 14. The Chronometer", *Motor Boating* July 1947, 49–51.

Pontani, Filippomaria, "The World on a Fingernail: an Unknown Byzantine Map, Planudes, and Ptolemy", *Traditio* 65 (2010) 177–200. https://www.jstor.org/stable/41417993 (accessed 6 November 2023).

Raby, Julian, "Mehmed the Conqueror's Greek Scriptorium", *Dumbarton Oaks Papers* 37 (1983) 15–34.

Ragep, F. Jamil, "New Light on Shams: the Islamic Side of Σὰμψ Πουχάρης", in *Politics, Patronage and the Transmission of Knowledge in 13th–15th Century Tabriz*, ed. Judith Pfeiffer, Leiden: Brill, 2014, pp. 231–247.

Rinner, Elisabeth, *Zur Genese der Ortskoordinaten Kleinasiens in der Geographie des Klaudios Ptolemaios*, Bern, Tilman Sauer, 2013.

Rossikopoulos, Dimitrios A., "The geodetic sciences in Byzantium", in *Measuring and Mapping the Earth*, eds. Aristeidis Fotiou, Ioannis Paraschakis, Dimitrios Rossikopoulos, Thessaloniki: Special issue for Professor Emeritus Christogeorgis Katsikis, 2015, pp. 1–20. https://www.topo.auth.gr/wp-content/uploads/sites/111/2021/12/01_Rossikopoulos.pdf (accessed 6 November 2023).

Russel, Jeffrey B., *Inventing the Flat Earth*, New York: Praeger Press. https://konstantinus.com/wp-content/uploads/2017/01/Russell_Flat_Earth.pdf (accessed 6 November 2023).

Saliba, George, *Islamic Science and the Making of the European Renaissance*, Cambridge: MIT Press, 2007.

Sandys, John E., *A History of Classical Scholarship: From the Sixth Century B.C. to the End of the Middle Ages*, vol. 1, Cambridge, UK: Cambridge University Press, 1903.

Shcheglov, Dmitry A., "Eratosthenes' Contribution to Ptolemy's Map of the World", *Imago Mundi* 69:2 (2017) 159–175. https://doi.org/10.1080/03085694.2017.1312112 (accessed 6 November 2023).

Strabo, *Geographica* (1st Century BCE). Edited and translated version: https://www.perseus.tufts.edu/hopper/text?doc=Perseus%3Atext%3A1999.01.0239 (accessed 6 November 2023).

Stückelberger, Alfred, "Planudes und die *Geographia* des Ptolemaios", *Museum Helveticum* 53 (1996) 179–205.

Tapan, Miraç, "Journey to the center of the world: Is it far?", *Daily Sabah*, 25 September 2018. https://www.dailysabah.com/feature/2018/09/25/journey-to-the-center-of-the-world-is-it-far (accessed 6 November 2023).

Tihon, Anne, "The Astronomy of George Gemistus Plethon", *Journal for the History of Astronomy* 29 (1998) 109–116. https://doi.org/10.1177/002182869802900203 (accessed 6 November 2023).

Tihon, Anne, "Les Sciences Exactes à Byzance", *Byzantion* 79 (2009) 380–434. https://www.jstor.org/stable/44173183 (accessed 6 November 2023).

Tozer, Henry F., *A History of Ancient Geography*, Cambridge: Cambridge University Press, 1897. https://wellcomecollection.org/works/jxqnp663 (accessed 6 November 2023).

Tupikova, Irina, Matthias Schemmel and Klaus Geus, *Travelling along the Silk Road: A new interpretation of Ptolemy's coordinates*, Preprint, Max-Planck-Institut für Wissenschaftsgeschichte, 2014. https://www.mpiwg-berlin.mpg.de/Preprints/P465.PDF (accessed 6 November 2023).

Vasiliev, Alexander A., *History of the Byzantine Empire, 324–1453* I, Madison: University of Wisconsin Press, 1958.

Wilson, Nigel G., *From Byzantium to Italy: Greek Studies in the Italian Renaissance*, Baltimore: Bloomsbury Publishing, 1992.